\documentclass[aps,prd,preprint,draft,showpacs]{revtex4}
\usepackage{graphicx,epsf,amssymb}
\newcommand{\be}{\begin{equation}} 
\newcommand{\ee}{\end{equation}}

\newcommand{\bea}{\begin{eqnarray}} 
\newcommand{\eea}{\end{eqnarray}} 
\newcommand{\Tr}{{\rm Tr}}

\newcommand{\NeqFour}{{\cal N} =4}

\def\CP{\mathbb{CP}}

\newif\ifdraft
\drafttrue
\newif\ifpreprint
\preprinttrue

\def\sect#1{section~{\ref{#1}}}
\def\app#1{appendix~{\ref{#1}}}

\def\fig#1{fig.~{\ref{#1}}}

\def\tree{{\rm tree}}

\def\Tr{\, {\rm Tr}}

\def\hf{{{1\over2}}}
\def\Neqfour{{\cal N}=4}
\def\Neqone{{\cal N}=1}
\def\NeqFour{{\cal N}=4}
\def\NeqOne{{\cal N}=1}

\def\spa#1.#2{\left\langle#1\,#2\right\rangle}
\def\spb#1.#2{\left[#1\,#2\right]}
\def\sand#1.#2.#3{%
\left\langle\smash{#1}{\vphantom1}^{-}\right|{#2}%
\left|\smash{#3}{\vphantom1}^{-}\right\rangle}
\def\sandp#1.#2.#3{%
\left\langle\smash{#1}{\vphantom1}^{-}\right|{#2}%
\left|\smash{#3}{\vphantom1}^{+}\right\rangle}
\def\sandpp#1.#2.#3{%
\left\langle\smash{#1}{\vphantom1}^{+}\right|{#2}%
\left|\smash{#3}{\vphantom1}^{+}\right\rangle}
\def\sandpm#1.#2.#3{%
\left\langle\smash{#1}{\vphantom1}^{+}\right|{#2}%
\left|\smash{#3}{\vphantom1}^{-}\right\rangle}
\def\sandmp#1.#2.#3{%
\left\langle\smash{#1}{\vphantom1}^{-}\right|{#2}%
\left|\smash{#3}{\vphantom1}^{+}\right\rangle}
\def\sandmm#1.#2.#3{%
\left\langle\smash{#1}{\vphantom1}^{-}\right|{#2}%
\left|\smash{#3}{\vphantom1}^{-}\right\rangle}
\def\spab#1.#2.#3{\sandmm#1.#2.#3}
\def\spba#1.#2.#3{\sandpp#1.#2.#3}
\def\spaa#1.#2.#3.#4{\sandmp#1.{#2#3}.#4}
\def\spbb#1.#2.#3.#4{\sandpm#1.{#2#3}.#4}

\newbox\charbox
\newbox\slabox
\def\s#1{{      
        \setbox\charbox=\hbox{$#1$}
        \setbox\slabox=\hbox{$/$}
        \dimen\charbox=\ht\slabox
        \advance\dimen\charbox by -\dp\slabox
        \advance\dimen\charbox by -\ht\charbox
        \advance\dimen\charbox by \dp\charbox
        \divide\dimen\charbox by 2
        \raise-\dimen\charbox\hbox to \wd\charbox{\hss/\hss}
        \llap{$#1$}
}}

\def\eqn#1{eq.~(\ref{#1})}
\def\Eqn#1{Equation~(\ref{#1})}
\def\eqns#1#2{eqs.~(\ref{#1}) and~(\ref{#2})}

\def\e{\epsilon}
\def\eps{\epsilon}

\def\Gr{{\rm Gr}}

\def\lr{\leftrightarrow}

\def\Li{\mathop{\rm Li}\nolimits}
\def\Split{\mathop{\rm Split}\nolimits}
\def\oneloop{{1 \mbox{-} \rm loop}}

\def\cg{\hat{c}_\Gamma}
\def\Ord{{\cal O}}
\def\mod{\mathop{\rm mod}\nolimits}

\def\sandp#1.#2.#3{%
\left\langle\smash{#1}{\vphantom1}^{+}\right|{#2}%
\left|\smash{#3}{\vphantom1}^{+}\right\rangle}

\def\Ksl{\s{K}}

\def\Fact{{\cal F}}

\def\Fs#1#2{F^{{#1}}_{n:#2}}

\def\Fhard{\Fs{{\rm 2m}\,h}}

\newbox\ourfigbox
\def\SizedFigureWithCaption#1#2#3{%
\setbox\ourfigbox
  \hbox{\hss\epsfxsize #1 \epsfbox{#2}\hss}
\hbox to \wd\ourfigbox{\vbox{\noindent\copy\ourfigbox\break
\vskip -6mm      \hbox to \wd\ourfigbox{\hss#3\hss}}}
}

\def\spa#1.#2{\left\langle#1\,#2\right\rangle}
\def\spb#1.#2{\left[#1\,#2\right]}
\def\lor#1.#2{\left(#1\,#2\right)}
\def\sand#1.#2.#3{%
\left\langle\smash{#1}{\vphantom1}^{-}\right|{#2}%
\left|\smash{#3}{\vphantom1}^{-}\right\rangle}
\def\sandpp#1.#2.#3{%
\left\langle\smash{#1}{\vphantom1}^{+}\right|{#2}%
\left|\smash{#3}{\vphantom1}^{+}\right\rangle}
\def\sandpm#1.#2.#3{%
\left\langle\smash{#1}{\vphantom1}^{+}\right|{#2}%
\left|\smash{#3}{\vphantom1}^{-}\right\rangle}
\def\sandmp#1.#2.#3{%
\left\langle\smash{#1}{\vphantom1}^{-}\right|{#2}%
\left|\smash{#3}{\vphantom1}^{+}\right\rangle}

\begin{document}
\hfuzz 10 pt


\ifpreprint
\noindent
UCLA/04/TEP/43
\hfill DFTT 26/2004
\hfill  DCPT/04/136
\hfill IPPP/04/68
\hfill SLAC--PUB--10810
\hfill Saclay/SPhT--T04/131
\hfill  NSF--KITP--04--114 
\fi

\title{All Non-Maximally-Helicity-Violating One-Loop \\
 Seven-Gluon Amplitudes in ${\cal N}=4$ Super-Yang-Mills Theory%
\footnote{Research supported in part by the US Department of 
 Energy under contracts DE--FG03--91ER40662 and DE--AC02--76SF00515,
 by the National Science Foundation under grants PHY99--07949, 
 and by the {\it Direction des Sciences de la Mati\`ere\/}
 of the {\it Commissariat \`a l'Energie Atomique\/} of France.}}

\author{Zvi Bern}
\affiliation{ Department of Physics and Astronomy, UCLA\\
\hbox{Los Angeles, CA 90095--1547, USA}
}

\author{Vittorio Del Duca} 
\affiliation{ \\
 Istituto Nazionale di Fisica Nucleare \\ 
             Sez.~di Torino \\
 \hbox{via P. Giuria, 1--10125 Torino, Italy}
}

\author{Lance J. Dixon} 
\affiliation{ Stanford Linear Accelerator Center \\ 
              Stanford University\\
             Stanford, CA 94309, USA\\
 and\\
Institute of Particle Physics Phenomenology\\
\hbox{Department of Physics, University of Durham} \\
Durham, DH1 3LE, UK
}

\author{David A. Kosower} 
\affiliation{Service de Physique Th\'eorique, CEA--Saclay\\ 
          F--91191 Gif-sur-Yvette cedex, France
}

\date{October 21, 2004}

\begin{abstract}
We compute the non-MHV one-loop seven-gluon amplitudes in ${\cal N}=4$
super-Yang-Mills theory, which contain three negative-helicity gluons
and four positive-helicity gluons.  There are four independent
color-ordered amplitudes, $({-}{-}{-}{+}{+}{+}{+})$,
$({-}{-}{+}{-}{+}{+}{+})$, $({-}{-}{+}{+}{-}{+}{+})$ and
$({-}{+}{-}{+}{-}{+}{+})$.  The MHV amplitudes containing two
negative-helicity and five positive-helicity gluons were computed
previously, so all independent one-loop seven-gluon helicity
amplitudes are now known for this theory.  We present partial
information about an infinite sequence of next-to-MHV one-loop
helicity amplitudes, with three negative-helicity and $n-3$
positive-helicity gluons, and the color ordering
$({-}{-}{-}{+}{+}\cdots{+}{+})$; we give a new coefficient of one
class of integral functions entering this amplitude.  We discuss the
twistor-space properties of the box-integral-function coefficients in
the amplitudes, which are quite simple and suggestive.
\end{abstract}

\pacs{11.15.Bt, 11.25.Db, 11.25.Tq, 11.55.Bq, 12.38.Bx \hspace{1cm}}

\maketitle



\renewcommand{\thefootnote}{\arabic{footnote}}
\setcounter{footnote}{0}


\section{Introduction}
\label{IntroSection}

Gauge theories play a central role in modern theoretical physics.
They form the backbone of the Standard Model of particle interactions.
They also play an increasingly important role in our understanding of
string theories via the AdS/CFT correspondence. That correspondence is
a strong--weak coupling duality between type~IIB string theory on an
${\rm AdS}_5 \times S^5$ background and four-dimensional $\NeqFour$
supersymmetric gauge theory. In a recent paper,
Witten~\cite{WittenTopologicalString} proposed a {\it weak--weak}
coupling duality between $\NeqFour$ supersymmetric gauge theory and
the topological open-string $B$ model on $\CP^{3|4}$ using
$D$-instanton contributions.  This proposal generalizes Nair's earlier
description~\cite{Nair} of the simplest gauge-theory amplitudes as
correlation functions on $\CP^{1}$.
Berkovits~\cite{Berkovits,BerkovitsMotl}, Neitzke and
Vafa~\cite{Vafa}, and Siegel~\cite{Siegel} have given alternative
descriptions of such a possible topological dual to the $\NeqFour$
theory.  Both of these dualities motivate the computation of
perturbative amplitudes in gauge theory.

Indeed, the availability of an extensive literature of explicit
results for both tree-level and loop amplitudes in gauge theories
has provided an important stimulus to and guide in building the
topological constructions.  Most notable are the series of
helicity amplitudes with arbitrarily many external gluons, 
the maximally helicity violating
(MHV) amplitudes at tree level~\cite{ParkeTaylor,Recurrence}, and
at one loop in supersymmetric theories~\cite{Neq4Oneloop,Neq1Oneloop}.  
These amplitudes display a remarkable simplicity. 
The entire amplitude at a given multiplicity is
much simpler than the typical Feynman diagram, chosen from the great
number that are assembled into the amplitude in a traditional approach.  
There are indications that this simplicity continues to higher
loops~\cite{BRY,IterationRelationA,IterationRelationB}.

The forthcoming generation of experiments probing beyond the Standard
Model at the LHC provides another important motivation for computing
amplitudes in perturbative gauge theory.  Precision measurements at SLC
and LEP have proven to be a powerful means of advancing our understanding of
the Standard Model.  A drive towards precision physics at hadron
colliders will require a corresponding theoretical effort in precision
calculations in QCD, to higher loops and multiplicities.  
While a traditional field-theoretic point of view would
view the pure glue contributions to QCD amplitudes as a small part of
the $\NeqFour$ result, it is
more natural to view matters the other way around.  The $\NeqFour$
results are much simpler than the QCD ones and can be regarded as a
building block for the latter~\cite{Neq4Oneloop}.  For example, the
gluon (or quark) circulating in the loop in a one-loop $n$-gluon QCD
amplitude can be decomposed into a linear combination of an ${\cal
N}=4$ super-multiplet (absent in the quark case), an ${\cal N} =1$
chiral super-multiplet, and a scalar in the loop.  Each piece has a
different analytic structure, so it makes sense to separate the
components.  The $\NeqFour$ amplitudes can always be written as a sum
over a more limited class of integral functions than the other
components, as they require only scalar box integrals.  These
four-point integrals have no loop momenta in the numerator of the
integral~\cite{Neq4Oneloop}.  The remaining computational problem is
to determine the coefficient with which each scalar box integral
appears in the amplitude.  This simplicity can be traced back to the
much-improved ultraviolet behavior of the $\NeqFour$ theory. At the
same time, the $\NeqFour$ component captures the leading infrared
singularities present in the QCD amplitude.

Calculations in QCD and in $\NeqFour$ super-Yang-Mills theory are
closely related to each other.  In general, an $\NeqFour$ amplitude
may be extracted from a corresponding QCD amplitude, if the
number of gluonic and fermionic states is tracked in the computation.  
The conversion of a QCD amplitude to an $\Neqfour$ super-Yang-Mills 
amplitude is accomplished
by assigning 8 spin degrees of freedom to gluons circulating in the
loops, and modifying the multiplicity of the fermions to also carry a
total of 8 spin degrees of freedom. The color algebra also needs to be
modified to account for the gluinos being in the adjoint color
representation, instead of the quarks' fundamental color representation. 
With these modifications, the theory has the number of degrees 
of freedom of ten-dimensional $\Neqone$ super-Yang-Mills theory
compactified to four dimensions, which is another name for
$\NeqFour$ super-Yang-Mills theory.
This simple conversion holds at any loop order.
As a non-trivial example, after carrying out this conversion on the
two-loop four-gluon amplitudes given in ref.~\cite{Twoloopgggg}, they
agree perfectly~\cite{IterationRelationA} with the previously 
determined $\NeqFour$ amplitudes~\cite{BRY}.  The analytical results
presented here can therefore serve as benchmarks for testing
algorithms~\cite{NumericalOneLoop}
for direct numerical evaluation of high-multiplicity one-loop 
amplitudes in QCD.

The experimentally-driven computations are technically complicated, so
they require more powerful methods than textbook ones.  In the past
decades, a number of new approaches have been developed to cope with
this complexity, including helicity methods~\cite{SpinorHelicity},
color decompositions~\cite{TreeColor,TreeReview,BKColor,DDDM},
recursion relations~\cite{Recurrence}, supersymmetry Ward
identities~\cite{SWI}, ideas based on string theory~\cite{BKgggg,CSW},
and the unitarity-based
method~\cite{Neq4Oneloop,Neq1Oneloop,UnitarityMachinery,LoopReview,
TwoloopSplit}.  The latter technique has been applied to numerous
calculations, most recently the two-loop calculation of all helicity
amplitudes for gluon--gluon scattering~\cite{AllPlusTwo} and the
two-loop splitting amplitudes $g \rightarrow gg$~\cite{TwoloopSplit}.
(The latter agree with the explicit collinear limits of certain
two-loop amplitudes~\cite{TwoloopSplit,GloverSplitting}.)

Forefront calculations still tax available technologies to their
utmost, which motivates the continuing development of new
techniques. The twistor-space representation, introduced along with the
topological string theories mentioned above, shows great promise as a
source of new methods for performing cutting-edge
calculations. Indeed, Cachazo, Svr\v{c}ek, and Witten~\cite{CSW} have
used observations about the twistor-space structure of amplitudes to
formulate a new and simple set of rules based on {\it MHV vertices}
(off-shell continuations of the Parke-Taylor MHV amplitudes) which can
be used to compute all tree-level amplitudes in unbroken gauge theory.
Their rules extend to processes with external fermions and scalars as
well as gluons~\cite{Khoze}, and lead to explicit results for
next-to-MHV amplitudes~\cite{CSW,Khoze,NMHVTree} and checks of
`googly' amplitudes (with two positive-helicity, and the rest
negative-helicity, gluons)~\cite{RSV1,OtherGoogly,WittenParity}.  The
rules can be implemented either analytically or numerically.  The
computational efficiency of these rules may be further enhanced with a
recursive rearrangement~\cite{MHVRecursive}.

A critical question is how to extend these ideas to loop calculations.
The challenge is to reduce loop calculations to a purely algebraic
problem of polynomial complexity, avoiding the exponential explosion
in the complexity of intermediate stages that would be encountered
with brute force tensor reduction and integration methods.  A direct
topological-string approach appears to lead to complications, mixing
in non-unitary states from conformal
supergravity~\cite{BerkovitsWitten}.  However, in an important step,
Brandhuber, Spence, and Travaglini~\cite{BST} have shown that one can
compute one-loop amplitudes using exactly the same MHV vertices that
work at tree level.  They showed explicitly how to compute the
infinite sequence of one-loop MHV amplitudes in $\NeqFour$
super-Yang-Mills theory.  Many technical steps in their computation
parallel the original cut-based method~\cite{Neq4Oneloop}; however,
the new computation is quite different conceptually, and probes the
off-shell continuation of the MHV vertices.  The BST calculation also
reaffirms the basic simplicity of the twistor-space structure of
one-loop amplitudes.  Taking into account~\cite{CSWIII} a `holomorphic
anomaly' brings an earlier investigation of this
structure~\cite{CSWII} into agreement with the picture emerging from
the BST calculation.  A delta function in the holomorphic anomaly for
a unitarity cut completely freezes the integration over the
intermediate phase space variables~\cite{BBKRTwistor,Cachazo}, making
it simple to evaluate the action of the differential operators for
twistor space co-linearity or co-planarity on the cuts of one-loop
amplitudes.  The anomaly has already been applied by
Cachazo~\cite{Cachazo}, in conjunction with unitarity, to derive
algebraic equations relating coefficients of scalar box integrals to
the rational functions making up a cut integrand.

In this paper we present new results for amplitudes in the $\NeqFour$
gauge theory.  We will give compact formul\ae{} for all four helicity
configurations required for the seven-point next-to-MHV (NMHV) amplitude.
These amplitudes were computed by evaluating unitarity cuts in
various channels, the same method used previously to obtain the
all-multiplicity MHV helicity amplitudes~\cite{Neq4Oneloop} and the
six-point NMHV helicity amplitudes~\cite{Neq1Oneloop}.  In this case,
however, the unitarity cuts are more complicated.  We have reduced them to
a standard set of (cut) scalar box integrals using integral reduction methods
implemented on the computer.  The coefficients of the scalar boxes in the
full amplitude are equal to the coefficients of the cut scalar boxes
that have a cut in the channel considered.  The resulting expressions for
the coefficients are analytical, but quite large.
However, remarkably simple expressions exist for these quantities.  The
simple expressions can be found by considering the analytic behavior of
the coefficients in various limiting regions of the seven-point phase
space, and using that information to build ans\"atze for the coefficients.
The ans\"atze can then be checked numerically with high precision against
the large expressions at random kinematic points.

There are a large number of box coefficients to determine
(92, after invoking some reflection symmetries).  However,
their structure can be fit into general patterns, which are simple to 
describe in twistor-space language.  The twistor-space structure of
the full amplitude, including the box integrals multiplying these
coefficients, is less transparent because of the holomorphic anomaly
which affects the integrals~\cite{CSWIII}.
On the other hand, the twistor-space properties of the coefficients square well
with the recent application of the holomorphic anomaly to computation
of ${\cal N}=4$ box coefficients~\cite{Cachazo}.

We also give an all-$n$ formula for the coefficient of a particular class
of (three-mass) box integrals in the adjacent-minus NMHV amplitude, 
the helicity configuration $({-}{-}{-}{+}{+}\cdots{+}{+})$.
We study the twistor-space structure of this part of
the amplitude and find that it is consistent with expectations of
simplicity emerging from the BST calculation and the holomorphic anomaly.  
These results should provide a useful guide to the analytic structure
likely to emerge in other scalar-box coefficients in this amplitude
and in other amplitudes.

As this paper was being completed, ref.~\cite{BCF} appeared, in which
one of the four seven-point helicity amplitudes presented here,
$({-}{-}{-}{+}{+}{+}{+})$, is computed via the holomorphic anomaly 
and unitarity.  Although some of the expressions obtained for the box
coefficients are more complicated than ours, we have compared them 
all numerically and find complete agreement.

This paper is organized as follows.  In section~\ref{ReviewSection}
we review the color and helicity decompositions of tree-level and one-loop
gauge amplitudes, as well as the structure of the MHV amplitudes
at these orders. 
In section~\ref{CalculationSection} we describe the application
of the unitarity-based technique to the seven-point NMHV computation,
including an outline of the integral reduction approach and its
implications for the denominators of different classes of box
coefficients.   We also sketch how the coefficients were simplified.
Section~\ref{ResultsSection} details the results for the
seven-point amplitudes by listing the independent box integral coefficients.
Section~\ref{ConsistencySection} describes consistency checks that 
were applied to the results.  These checks include the behavior as
two gluons $i$ and $j$ become collinear, {\it i.e.} as the kinematic 
invariant $s_{ij} = (k_i+k_j)^2 = 2k_i\cdot k_j \to 0$.   
They also include the behavior as multi-particle invariants vanish; that is,  
$s_{ijk} = (k_i+k_j+k_k)^2 \to 0$.
In section~\ref{AllNSection} we give one of the box integral coefficients
in the all-$n$ NMHV helicity amplitude $({-}{-}{-}{+}{+}\cdots{+}{+})$.
Section~\ref{TwistorSection} analyzes the twistor-space properties
of the seven-point box coefficients, and of the term in the \hbox{all-$n$}
NMHV helicity amplitude from section~\ref{AllNSection}.
In section~\ref{ConclusionsSection} we present our conclusions.
There are two appendices.  Appendix~\ref{IntegralsAppendix} collects
the dimensionally-regulated scalar box integral functions appearing in 
the $\NeqFour$ amplitudes.  Appendix~\ref{FactorizationAppendix}
describes an example of how one of the seven-point helicity amplitudes 
can be factorized onto a multi-particle pole.


\section{Review of Previous Amplitude Results}
\label{ReviewSection}

We now briefly summarize results for previously computed
amplitudes in $\NeqFour$ super-Yang-Mills theory.
At tree level it is convenient to use the color
decomposition~\cite{TreeColor,TreeReview} of amplitudes
\begin{equation}
{\cal A}_n^\tree(\{k_i,\lambda_i,a_i\}) = 
\sum_{\sigma \in S_n/Z_n} \Tr(T^{a_{\sigma(1)}}\cdots T^{a_{\sigma(n)}})\,
A_n^\tree(\sigma(1^{\lambda_1},\ldots,n^{\lambda_n}))\,,
\label{TreeColorDecomposition}
\end{equation}
where $S_n/Z_n$ is the group of non-cyclic permutations on $n$
symbols, and $j^{\lambda_j}$ denotes the $j$-th momentum and helicity
$\lambda_j$.  The $T^a$ are fundamental
representation SU$(N_c)$ color matrices normalized so that
$\Tr(T^a T^b) = \delta^{ab}$.  The color-ordered amplitude $A_n^\tree$ 
is invariant under a cyclic permutation of its arguments.

We describe the amplitudes using the spinor helicity formalism.
In this formalism amplitudes are expressed in terms of spinor
inner-products,
\begin{equation}
\spa{j}.{l} = \langle j^- | l^+ \rangle = \bar{u}_-(k_j) u_+(k_l)\,, 
\hskip 2 cm
\spb{j}.{l} = \langle j^+ | l^- \rangle = \bar{u}_+(k_j) u_-(k_l)\, ,
\end{equation}
where $u_\pm(k)$ is a massless Weyl spinor with momentum $k$ and plus
or minus chirality~\cite{SpinorHelicity,TreeReview}. Our convention
is that all legs are outgoing. The notation used here follows the
standard QCD literature, with
\begin{equation}
\spa{i}.{j} \spb{j}.{i} = 2 k_i \cdot k_j = s_{ij}\,.
\end{equation}
(Note that the square bracket $\spb{i}.{j}$ differs by an overall sign
compared to the notation commonly used in twistor-space
studies~\cite{WittenTopologicalString}.)  For non-MHV amplitudes we
also use the abbreviated notation,
\begin{eqnarray}
 \spba{i}.{(a+b)}.j  &=& \spba{i}.{(\s{k}_a+\s{k}_b)}.{j}\,,\nonumber\\
 \spaa{i}.{(a+b)}.{(c+d)}.j &=& 
   \spaa{i}.{(\s{k}_a+\s{k}_b)}.{(\s{k}_c+\s{k}_d)}.j 
    \,.
\end{eqnarray}

We denote the sums of cyclicly-consecutive external momenta by
\begin{equation}
K^\mu_{i\ldots j} \equiv (k_i + k_{i+1} + \cdots + k_{j-1} + k_j)^\mu \,,
\label{KDef}
\end{equation}
where all indices are mod $n$ for an $n$-gluon amplitude.
The invariant mass of this vector is $s_{i\ldots j} = K_{i\ldots j}^2$.
In the seven-point case, using momentum conservation we just need to
consider two- and three-particle invariant masses, which are denoted by
\begin{equation}
s_{ij} \equiv (k_i+k_j)^2 = 2k_i\cdot k_j,
\qquad \quad
s_{ijk} \equiv (k_i+k_j+k_k)^2.
\label{TwoThreeMassInvariants}
\end{equation}
In color-ordered amplitudes only invariants with cyclicly-consecutive
arguments appear, $s_{i,i+1}$ and $s_{i,i+1,i+2}$.

The simplest of the partial amplitudes are the maximally
helicity-violating (MHV) Parke-Taylor tree amplitudes~\cite{ParkeTaylor}
with two negative-helicity gluons and the rest of positive helicity,
\begin{equation}
  A_{jk}^{\tree \rm\ MHV}(1,2,\ldots,n)
 =  i\, { {\spa{j}.{k}}^4 \over \spa1.2\spa2.3\cdots\spa{n}.1 }\, ,
\label{PT}
\end{equation}
where $j$ and $k$ label the negative-helicity legs.

For one-loop amplitudes, the color decomposition is
similar~\cite{BKColor}.  When all internal particles transform in
the adjoint representation of SU$(N_c)$, as is the case for 
$\NeqFour$ supersymmetric Yang-Mills theory, we have
\begin{equation}
{\cal A}_n^\oneloop ( \{k_i,\lambda_i,a_i\} ) =
  \sum_{c=1}^{\lfloor{n/2}\rfloor+1}
      \sum_{\sigma \in S_n/S_{n;c}}
     \Gr_{n;c}( \sigma ) \,A_{n;c}(\sigma) \,,
\label{ColorDecomposition}
\end{equation}
where ${\lfloor{x}\rfloor}$ is the largest integer less than or equal to $x$.
The leading color-structure factor
\begin{equation}
\Gr_{n;1}(1) = N_c\ \Tr (T^{a_1}\cdots T^{a_n} ) \,, 
\end{equation}
is $N_c$ times the tree color factor.  The subleading color
structures are given by
\begin{equation}
\Gr_{n;c}(1) = \Tr ( T^{a_1}\cdots T^{a_{c-1}} )\,
\Tr ( T^{a_c}\cdots T^{a_n}).
\end{equation}
$S_n$ is the set of all permutations of $n$ objects,
and $S_{n;c}$ is the subset leaving $\Gr_{n;c}$ invariant.

The one-loop subleading-color partial amplitudes are given by a sum over
permutations of the leading-color ones~\cite{Neq4Oneloop},
\begin{equation}
 A_{n;c}(1,2,\ldots,c-1;c,c+1,\ldots,n)\ =\
 (-1)^{c-1} \sum_{\sigma\in COP\{\alpha\}\{\beta\}} A_{n;1}(\sigma) \,,
\label{SublAnswer}
\end{equation}
where $\alpha_i \in \{\alpha\} \equiv \{c-1,c-2,\ldots,2,1\}$,
$\beta_i \in \{\beta\} \equiv \{c,c+1,\ldots,n-1,n\}$, and
$COP\{\alpha\}\{\beta\}$ is the set of all permutations of
$\{1,2,\ldots,n\}$ with $n$ held fixed that preserve the cyclic
ordering of the $\alpha_i$ within $\{\alpha\}$ and of the $\beta_i$
within $\{\beta\}$, while allowing for all possible relative orderings
of the $\alpha_i$ with respect to the $\beta_i$.  To obtain the
full amplitude, therefore, we need only compute the leading-color 
single-trace partial amplitudes.  The simple relation~(\ref{SublAnswer}) 
between leading- and subleading-color contributions is special to one loop;
at higher loops, new non-planar structures enter the subleading-color 
partial amplitudes.

We also collect here the results for the MHV partial amplitudes
in the $\NeqFour$ theory~\cite{Neq4Oneloop}.  These amplitudes
appear in various kinematic limits of the NMHV amplitudes, so
it is useful to understand their structure.
The MHV amplitudes are a simple linear combination of 
certain box integral functions,
\begin{equation}
 A_{n;1}^{\NeqFour\ {\rm MHV}} =
   c_\Gamma \, A_{n}^{\rm tree} \times  V_n^g \,.
\label{MHVAmplitudes}
\end{equation}
The factor $V_n^g$ ($n\ge 5$) depends on whether $n$ is odd ($n=2m+1$) 
or even ($n=2m$),
\begin{eqnarray}
(\mu^2)^{-\eps} V_{2m+1}^g &= &  \sum_{r=2}^{m-1} \sum_{i=1}^{n}
 F^{2{\rm m} \, e}(s_{i\ldots (i+r)}, s_{(i-1)\ldots (i+r-1)}, 
                   s_{i \ldots (i+r-1)}, s_{(i+r+1) \ldots (i-2)}) \nonumber \\
&& \hskip .2 cm \null
     + \sum_{i=1}^{n} F^{1{\rm m}} (s_{i-3,i-2}, s_{i-2,i-1}, 
                s_{i \ldots (i-4)})
       \,, \nonumber \\
(\mu^2)^{-\eps} V_{2m}^g &= & 
\sum_{r=2}^{m-2} \sum_{i=1}^{n} 
  F^{2{\rm m} \, e}(s_{i\ldots (i+r)}, s_{(i-1)\ldots (i+r-1)}, 
                   s_{i \ldots (i+r-1)}, s_{(i+r+1) \ldots (i-2)}) \nonumber \\
&& \hskip .2 cm  \null 
+ \sum_{i=1}^{n}  F^{1{\rm m}} (s_{i-3,i-2}, s_{i-2,i-1}, 
                s_{i \ldots (i-4)}) \nonumber \\
&& \hskip .2 cm  \null 
+ \sum_{i=1}^{n/2} F^{2{\rm m} \, e}( s_{i \ldots (i+m-1)}, 
                                  s_{(i-1) \ldots (i+m-2)},
                                  s_{i \ldots (i+m-2)}, 
                                  s_{(i+m) \ldots (i-2)}) \,. \hskip 1 cm 
\label{CoeffDefine}
\end{eqnarray}
The box integral functions $F$ are defined in 
\app{IntegralsAppendix}.   They are essentially scalar box 
integrals, but multiplied by convenient normalization factors in order 
to remove all power-law behavior in the kinematic invariants, 
leaving only logarithmic behavior.  By dimensional analysis, 
the scale $\mu$ enters all dimensionally-regulated (unrenormalized) 
one-loop amplitudes as an overall factor of $(\mu^2)^\e$.

The four-point and five-point amplitudes appear in the multi-particle 
factorization limits of the seven-point amplitudes, as
discussed in \sect{ConsistencySection} and \app{FactorizationAppendix}.
The $n=4$ case is given by
\begin{equation}
A_{4;1}^{\NeqFour} = 2 \, c_\Gamma\,  (\mu^2)^{\eps}\, 
A^{\tree}_4  \times F^{{\rm 0m}}(s_{12}, s_{23})\,,
\label{FourPointAmplitude}
\end{equation}
for all non-vanishing helicity choices.
For $n=5$, the expression~(\ref{CoeffDefine}) reduces to
\begin{eqnarray}
A_{5;1}^{\NeqFour} &=& c_\Gamma\,  (\mu^2)^{\eps}\, 
A^{\tree}_5  \Bigl[ F^{{\rm 1m}}(s_{12}, s_{23}, s_{45})
                  + F^{{\rm 1m}}(s_{23}, s_{34}, s_{51})
                  + F^{{\rm 1m}}(s_{34}, s_{45}, s_{12})
\nonumber \\ && \hskip2.2cm 
                  + F^{{\rm 1m}}(s_{45}, s_{51}, s_{23})
                  + F^{{\rm 1m}}(s_{51}, s_{12}, s_{34})
            \Bigr] \,.
\label{FivePointAmplitude}
\end{eqnarray}

Besides the $\NeqFour$ MHV amplitudes, a number of other infinite series
of one-loop amplitude have been computed.  The $n$-point MHV gluon
amplitudes with an $\NeqOne$ chiral multiplet in the loop were also
computed using the unitarity method.  In QCD one-loop $n$-point amplitudes
with identical helicities are also known~\cite{AllPlus}.  At two loops
less is known.  A conjecture for the planar two-loop MHV amplitudes in
terms of one-loop amplitudes was presented in
ref.~\cite{IterationRelationA}, suggesting that the simplicity uncovered
at one loop persists to higher loops in the planar, leading-color limit
$N_c \to \infty$.

The one-loop non-MHV six-point amplitudes in $\NeqFour$ super-Yang-Mills
theory were computed in ref.~\cite{Neq1Oneloop}.  Here we
will compute all non-MHV seven-point amplitudes.  (For $n=6$ and $n=7$,
`non-MHV' coincides with `next-to-MHV'.) 
We expect that these new amplitudes will be helpful for unraveling 
the full $n$-point twistor-space structure. Indeed, their structure 
provided important guidance for obtaining the coefficients of a class 
of three-mass box integrals appearing in the all-$n$ next-to-MHV 
amplitudes, which we present in~\sect{AllNSection}.


\section{Calculational Approach}
\label{CalculationSection}

It seems clear that calculating the seven-point amplitude by
computing the 227,585 contributing Feynman diagrams one by one
is not the best way to proceed.  And an all-multiplicity result is 
simply not accessible to this standard approach.

Fortunately, the unitarity-based technique is better for this application.
(For a recent review, see ref.~\cite{TwoloopSplit}.)
The basic idea is to reconstruct a one-loop amplitude from its cuts 
or absorptive pieces, which are products of tree amplitudes.  
In general, one must calculate these cuts in $D$ dimensions. 
In this case there are no ambiguities in reconstructing amplitudes
in any massless theory.  
In the special case of supersymmetric theories,
one can even evaluate the cuts `in four dimensions' --- that is, by
assigning four-dimensional helicities to the states crossing the cut 
--- without encountering any ambiguities, for all terms in the amplitude
which survive as $\e \to 0$~\cite{Neq1Oneloop}.  
This property reflects the better ultraviolet behavior
of these theories.  Then the one-loop cuts reduce to products of 
tree-level helicity amplitudes.
Starting from tree amplitudes rather than diagrams
means that the extensive cancellations that occur in gauge theories
are taken into account before any loop integrations are done, which greatly
reduces the complexity of the calculations.  Because infinite
series of tree-level amplitudes are known, it also opens the door
to calculating infinite series of one-loop amplitudes.

The reconstruction is done by identifying the
corresponding Feynman integrals whose cuts yield the original integrands.
This is easier than doing the dispersion integrals explicitly, because
we take advantage of an extensive machinery for simplifying and computing
such integrals.

In calculating an $n$-point amplitude in gauge theory, one expects
to encounter $n$-point tensor integrals with tensor rank up to $n$.
In supersymmetric theories, the maximal tensor rank is lower,
a reflection of the better ultraviolet behavior of the 
theory~\cite{Neq1Oneloop}.  In ${\cal N}=4$ theories, the maximal
tensor rank is reduced to $n-4$~\cite{Superspace,Neq4Oneloop}.  
In addition, thorough use of the spinor-helicity representation of external
gluon polarization vectors can reduce both
the tensor rank and multiplicity of integrals we encounter upon
reconstructing them from the cuts.

Nonetheless, we do have to treat some higher-point and tensor integrals.
At one loop, on general grounds, one can reduce any loop integral in
$D$ dimensions to a combination of box integrals, triangles, and bubbles.
In ${\cal N}=4$ supersymmetric theories, only scalar box integrals are
needed, and no triangles or bubbles.  
The tensor higher-point integrals encountered, where loop momenta appear
in the numerator of the integrand, could be reduced to
scalar ones via brute-force Brown--Feynman/Passarino--Veltman
techniques~\cite{BFPV}, or alternatively via Feynman-parameter methods.
However, we have found it useful to follow a method adapted from
that originally proposed by van~Neerven and Vermaseren~\cite{vNV}, relying
on the fact that any vector in four dimensions can be written
in terms of a basis of four vectors. For example, we may expand
the vector $\ell_{[4]}^\mu$ as
\begin{equation}
\eps^{p_1p_2p_3p_4} \times \ell^\mu_{[4]}
=  \ell_{[4]}\cdot p_1 \ \eps^{\mu p_2p_3p_4}
 + \ell_{[4]}\cdot p_2 \ \eps^{p_1\mu p_3p_4}
 + \ell_{[4]}\cdot p_3 \ \eps^{p_1p_2\mu p_4}
 + \ell_{[4]}\cdot p_4 \ \eps^{p_1p_2p_3\mu } \,,
\label{LoopMomExpand}
\end{equation}
where $\eps^{\mu p_2p_3p_4}$ is shorthand for 
$\eps^{\mu}_{\phantom{\mu}\nu\sigma\rho} p_2^\nu p_3^\sigma p_4^\rho$,
and $\eps_{\mu\nu\sigma\rho}$ is the Levi-Civita tensor.
Of course, we are working
with dimensionally-regulated integrals, but we can split the loop
momentum vector $\ell^\mu$ into two parts, the four-dimensional components 
and the $(-2\e)$-dimensional ones.  The four-dimensional components can
be expanded according to \eqn{LoopMomExpand}, 
where the $p_i$ will be chosen to be either external momenta or sums of
external momenta (see below). 
In this case, we can take advantage of the fact that
\begin{eqnarray}
\ell_{[4]}\cdot p_i & = &
 {1\over2} (\ell_{[4]}^2 - (\ell_{[4]} - p_i)^2 + p_i^2 ) \nonumber \\
&=&  {1\over2} (\ell^2 - (\ell - p_i)^2 + p_i^2 ) \,,
\label{PropCancel}
\end{eqnarray}
in order to cancel propagators, or reduce the tensor rank of the integral.
In the unitarity-based method, $\ell^2$ is taken to be zero,
and so when we are using the spinor-helicity representation, what
appears in the numerator are only dot products of the loop momentum,
and spinor `sandwiches' of the loop momentum between spinors 
representing external momenta.  These objects depend only on the 
four-dimensional components of the loop momentum.
(Because we are evaluating the cuts in four dimensions, we have
dropped some terms in the cut integrand which are proportional to 
the square of the $(-2\e)$-dimensional components.  These pieces
lead only to $\Ord(\e)$-suppressed terms in the amplitudes, 
for supersymmetric theories.)

We do not choose a fixed basis of external vectors for all expansions.
For terms containing multiple factors of the loop momentum in the
numerator, we proceed in stages, expanding the successive occurrences of 
the loop momentum one by one.  At each stage, 
we choose the basis set for the expansion to consist
of external momenta, not of the amplitude, but of the subject integral. 
In particular, after the first stage of this procedure, 
one of the external legs will consist of a {\it sum} of the original 
external momenta, because of the cancellation of one of the propagators
using \eqn{PropCancel}.
We choose such a sum to be an element of the basis set for the second stage.
Using the new basis set in the second stage ensures that the 
loop-momentum-containing quantities $\ell^2$ and
$(\ell-p_i)^2$ appearing on the right-hand side of \eqn{PropCancel} 
will continue to cancel some propagator in the integral at this stage.
We proceed similarly in the further stages.
We call this approach a `pivoting' technique.
Since dot products of the loop momentum with external legs of any one-loop 
integral are always expressible in terms of inverse propagators and 
external invariants, as in \eqn{PropCancel},
our choice always allows a given integral to be reduced
into a sum of lower-rank and lower-point integrals.
(Were we to have chosen a fixed external basis set, expansions beyond
the first would {\it not\/} necessarily be expressible in terms
of inverse propagators of the daughter integrals, complicating the
analysis.)

In principle the pivoting technique can be applied to any cut
(or even to full uncut diagrams if desired).  For the $\NeqFour$ 
application, however, it is only necessary to consider cuts
in three-particle channels, those with invariants $s_{i,i+1,i+2}$
crossing the cut.   That is because box integrals contain cuts
in several channels, and for the seven-point process
at least one channel is always a three-particle channel.
Such a cut is the product of a five-point tree amplitude
and a six-point tree amplitude, or it may be a sum of
such terms, the sum corresponding to multiple members of the
$\NeqFour$ supermultiplet crossing the cut.   It is useful
to divide the cuts into `singlet' and `nonsinglet' pieces.
The singlet pieces have a total helicity of $\pm2$ crossing the cut.
They can only come from gluonic intermediate states.  
The nonsinglet pieces have a total helicity of $0$ crossing the cut,
and receive contributions from the entire $\NeqFour$ supermultiplet,
gluons, fermions and scalars.   One can further divide the singlet and
nonsinglet pieces according to whether the six-point tree amplitude(s)
are MHV or non-MHV.  So there are four types of cut building blocks:
\begin{itemize}
\item {\it singlet-MHV}.  This cut is the simplest type to evaluate, 
and can be done directly in a compact form for all $n$; 
see ref.~\cite{Cachazo} and \sect{AllNSection}).
\item {\it nonsinglet-MHV}.  This cut is also simple; the sum over
$\NeqFour$ states can be performed as in ref.~\cite{Neq1Oneloop},
resulting in an expression which is identical to the singlet-MHV
case, except for overall, loop-momentum-independent, prefactors.
\item {\it singlet-non-MHV}.  This cut is more complicated, but
still only the product of two helicity amplitudes.
\item {\it nonsinglet-non-MHV}.  This cut is the most complicated,
but only needed to be evaluated as a cross-check.
\end{itemize}

\begin{figure}[t]
\centerline{\epsfxsize 2.25 truein \epsfbox{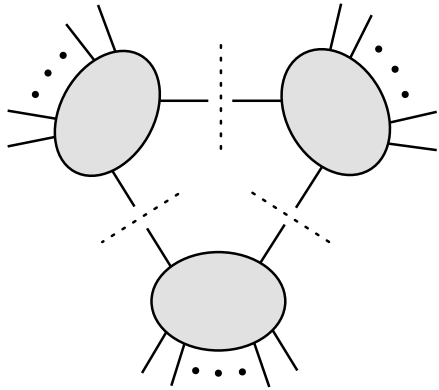}}
\caption[a]{\small A generalized `triple cut'.  The three
propagators cut by the dashed lines are required to be `open'.}
\label{triplecutFigure}
\end{figure}

It is also possible to consider {\it generalized} cuts which provide
very useful additional information about loop 
amplitudes~\cite{ZFourPartons,TwoloopSplit}. 
The example of a `triple cut' is shown in \fig{triplecutFigure}.
It is not a cut in the traditional sense of having a phase-space
integral which generates the imaginary part in a single channel.
It corresponds to a collection of underlying Feynman diagrams which
have three propagators `open', or uncancelled, in the same way that
an ordinary cut requires two propagators to be open.  Terms which
vanish as the `cut' propagators go on shell may be ignored.
The contributing terms can be represented as the product of
three tree amplitudes.
We will see in \sect{AllNSection} that analysis of such cuts
provides additional information, often quite easily.

Application of the pivoting technique to each type of cut
yields an expression in terms
of external spinor invariants, and scalar integrals up to $n$ points.
To reduce the six- and seven-point scalar integrals we used 
the formul\ae{} from appendix VI of ref.~\cite{Integrals5}.  To reduce
the five-point scalar integrals we made use of two different reduction
formul\ae{}.  The first formula is also from ref.~\cite{Integrals5} in 
the equivalent form from ref.~\cite{IntegralsN}, 
\begin{equation}
I_5 = {1\over2} \sum_{j=1}^n c_j I_4^{(j)} + \e \, c_0 \, I_5^{D=6-2\e},
\label{PentagonReduction}
\end{equation}
where $I_4^{(j)}$ is the daughter box integral derived from the
pentagon by omitting (the negative of) the propagator between 
legs $(j-1)$ and $j$; and where
\begin{eqnarray}
&&c_j = \sum_{l=1}^n S_{jl}^{-1} \,, 
\label{PentBoxCoeffs} \\
&&c_0 = \sum_{l=1}^n c_l \,,
\label{PentPentCoeff}
\end{eqnarray}
in terms of 
$S_{ij} = S_{ji} = - K_{i\ldots (j-1)}^2/2$ (and $S_{ii} = 0$).
The pentagon integral in $6-2\e$ dimensions, $I_5^{D=6-2\e}$, is
finite as $\e \to 0$, and it has a manifest prefactor of $\e$ in
\eqn{PentagonReduction}, so its contribution may be neglected
to the order we are working.

These reductions are valid in dimensional regularization.  The six-
and seven-point reduction formul\ae{} are equivalent to those of
Melrose~\cite{Melrose}, and van Neerven and Vermaseren~\cite{vNV}.
The five-point reduction formula is equivalent after dropping
the $\Ord(\e)$ terms.  The five-point reduction 
formula~(\ref{PentagonReduction}) is
valid for integrals, but does not hold point-by-point for
the integrands.  We found it convenient for certain parts of the 
calculation to use a reduction formula which is valid point-by-point.
We can obtain such a formula by adding appropriate terms containing
the Levi-Civita tensor and linear in the loop momentum, 
$\epsilon^{\mu\ldots} \ell_\mu \ldots$, which vanish upon performing the 
loop integration. The advantage of including these terms in the 
reduction is that one can check all algebra numerically, directly
on the integrand.

Once we have reduced the integrand to a sum of four-point integrands,
we are done, because amplitudes in the maximally supersymmetric theory
can be expressed in terms of scalar boxes, with no triangle or bubble
integrals.  (Due to $\NeqFour$ supersymmetry, each term in the cut 
integrand with $r-2$ non-cut propagators starts out with no more than 
degree $r-4$ tensor integrals.  Given this starting point, the pivoting 
reduction ensures that tensor boxes never appear.)

This procedure gives us the coefficients of the standard box integral
functions.  The forms emerging directly from the reductions are generally quite
complicated.  However, we can simplify them further, as discussed below.  
Somewhat surprisingly, many of them have a remarkably
simple form.  Indeed, the most `complicated' integrals that appear, the
three-mass boxes, have single-term coefficients, and the most complicated
coefficients have only four terms.  

The coefficients have various momentum-space singularities.
Some of the coefficient singularities are also singularities of the full
one-loop amplitude, such as the collinear singularities or poles in
multi-particle invariants $s_{ijk}$.  Other singularities are spurious.
These are introduced by the reduction procedure.  Some of these
singularities disappear coefficient by coefficient; others survive.  They
will cancel in the amplitude as a whole, as we will discuss in
section~\ref{ConsistencySection}; but this cancellation involves the box
integrals and their behavior in an essential way.  That is, each
coefficient can (and does in general) possess singularities not present in
the amplitude as a whole.  The pentagon
reduction~(\ref{PentagonReduction}), in particular, gives rise to
denominators which induce such singularities.  They can be read off from
\eqn{PentBoxCoeffs} by inspecting the determinant of the appropriate
pentagon $S_{ij}$ matrix.  In the seven-point calculation, we encounter
one-mass and two-mass pentagons.  The former can be treated as a special
case of the latter.  The non-adjacent two-mass pentagon (with external
legs $(k_1,k_2,P,k_3,Q)$, capital letters denoting massive legs) gives
rise to the following denominator,
\begin{equation}
2 k_1\cdot k_2 ( (P+k_2)^2 (P+k_3)^2 - P^2 (P+k_2+k_3)^2)
               ( (Q+k_3)^2 (Q+k_1)^2 - Q^2 (Q+k_1+k_3)^2).
\end{equation}
Each of the last two factors can be further factorized using the
spinor identity~\cite{ZFourPartons},
\def\Psl{\s{P}}
\begin{equation}
(P+k_1)^2 (P+k_2)^2 - P^2 (P+k_1+k_2)^2 = 
\langle 1^- | \Psl | 2^- \rangle \langle 2^- | \Psl | 1^- \rangle \,.
\label{PlanarDefinition}
\end{equation}
This type of denominator thus develops singularities when $P$
lies in the (space-time) plane spanned by $k_1$ and $k_2$,
\begin{equation}
P^\mu = a_1 k_1^\mu + a_2 k_2^\mu\,,
\end{equation}
for arbitrary values of $a_1$ and $a_2$.
For this reason, we call the factor in \eqn{PlanarDefinition} 
and the corresponding singularity a `planar' or `back-to-back' one.

The adjacent-mass two-mass pentagon (with external legs $(k_1,k_2,k_3,P,Q)$)
gives rise to the denominator,
\begin{equation}
4 k_1\cdot k_2\,k_2\cdot k_3\,
\Bigl[ (k_1+k_2+k_3)^2 (Q+k_1)^2 (P+k_3)^2 - 2 k_1\cdot k_2\, P^2 (Q+k_1)^2
   -2k_2\cdot k_3\,Q^2 (P+k_3)^2 \Bigr].
\end{equation}
The last factor, in brackets, also has a spinor factorization
(for $k_1 + k_2 + k_3 + P + Q = 0$),
\def\Qsl{\s{Q}}
\begin{equation}
\langle 1^- | \Qsl\Psl | 3^+ \rangle 
\langle 3^+ | \Psl\Qsl | 1^- \rangle\,.
\end{equation}
It therefore vanishes whenever
\begin{equation}
P = a_1 k_1 + a_2 k_2 + a_3 k_3\,,
\end{equation}
with 
\begin{equation}
a_2 = -a_1 { k_1\cdot k_2 + (1+a_3) k_1\cdot k_3 \over
            a_1 k_1\cdot k_2 + (1+a_3) k_2\cdot k_3 } \,.
\end{equation}
To distinguish it from the other kind of denominator and singularity, we will
call this type `cubic'.

In computing the all-$n$ coefficient discussed in \sect{AllNSection}, one
also encounters denominators arising from the reduction of certain
three-mass pentagon integrals.  It turns out that they also can
be expressed in terms of these sorts of `cubic' factors.

Knowledge of all of the singularities of the box coefficients
allows one to simplify them `numerically'.  The basic idea
is to generate kinematic points numerically which are close
to the various singular regions:  nearly collinear kinematics for 
each of the seven color-adjacent pairs ($i$, $i+1$);
kinematic points where each of the seven multi-particle invariants
$s_{i,i+1,i+2}$ almost vanish; and points near the relevant
`planar' and `cubic' singularities.  By studying the behavior
of the large analytical expressions for the box coefficients
in these regions, one can write down all the allowed factors
in the denominator of the expression.   In the collinear limit
where $k_i$ is nearly parallel to $k_{i+1}$, it is helpful
to be able to rotate $k_i$ and $k_{i+1}$ about their common sum.
The complex phase behavior under this rotation reveals whether
the denominator contains $\spa{i}.{(i+1)}$ or $\spb{i}.{(i+1)}$
(or both).  

The numerator of the box coefficient can often be completely
determined by the spinor homogeneity properties, which dictate how many
net powers of $\langle i^-|$ {\it vs.} $\langle i^+|$ occur,
in terms of the helicity of leg $i$.
In a few cases, we solved for a coefficient by writing down a sum 
of all possible expressions of the right dimension and spinor homogeneity,
and comparing it numerically to the actual expression at a number
of randomly-generated phase-space points.

The easy-two-mass box coefficients are the simplest to determine
in this way, because they cannot have any `cubic' singularities,
and because they turn out to be composed of just one term.
(The easy-two-mass box integral has a massless leg interposed between 
each massive one, which prevents it from being obtained from an 
adjacent-mass pentagon by cancelling a propagator.  The adjacent-mass
pentagons are the only source of cubic singularities.)
For example, in examining the coefficient $c_{147}$ for
$A_{7;1}^{\NeqFour}(1^-,2^-,3^-,4^+,5^+,6^+,7^+)$, one
quickly finds that it vanishes strongly as legs 2 and 3
become collinear, and as legs 6 and 7 become collinear.
The factors of ${\spa2.3}^3 {\spb6.7}^3$ in the numerator
of \eqn{mmmppppc147} can be established in this way.
The denominator factors in \eqn{mmmppppc147} are all easily obtained 
numerically as well, except for the question of whether 
$\spba1.{(6+7)}.5$ or its complex conjugate $\spab1.{(6+7)}.5$ correctly
describes the `planar' singularity, and similarly for 
$\spba6.{(7+1)}.2$ {\it vs.} $\spab6.{(7+1)}.2$.  
However, only one of these four possibilities can give the right spinor 
homogeneity for legs 1, 2, 5 and 6.   (The possibility of additional 
factors in the numerator is excluded in this case by dimensional analysis.)

In addition to simplifying the coefficients obtained from a direct 
calculation, we can also employ another tool.
The structure of infrared singularities provides equations which
can be used as consistency checks, as we shall discuss in
\sect{ConsistencySection}, or alternatively to solve for some of the 
coefficients.  The infrared singularities in the amplitude are
known on general grounds~\cite{UniversalIR} to be,
\def\cg{c_\Gamma}
\begin{equation}
A_{n;1}^{\NeqFour} \Bigr|_{\e\ {\rm pole}} =  
-{\cg\over\e^2} \sum_{i=1}^n \biggr({\mu^2\over -s_{i,i+1}}\biggl)^{\e}
\times A_n^{\tree}.
\label{OneLoopIRPoles}
\end{equation}
That is, the $1/\e$ poles are proportional to the seven-gluon 
NMHV tree amplitude~\cite{BGK} and
contain only logarithms of nearest-neighbor two-particle invariants.  
This implies that the coefficient of any given $\ln(-s_{i,i+1})/\e$ 
must be equal to the tree; 
and the coefficient of any $\ln(-s_{i,i+1,i+2})/\e$ must vanish.
On the other hand, the scalar box functions whose coefficients
we are computing contain both of these sorts of terms, and so both
types of equations are non-trivial.  Each box function in
eqs.~(\ref{Fboxes3m})--(\ref{Fboxes1m}) contains various
$\ln(-s_{i,i+1})/\e$ and $\ln(-s_{i,i+1,i+2})/\e$ terms with coefficients
0, $\pm1$ or $\pm{1\over2}$.   Thus the constraints arising 
from~\eqn{OneLoopIRPoles} become simple linear relations among the
coefficients, some of which involve the tree amplitude.
(Other linear relations may be determined from numerical evaluation
of the box coefficients.)
We use the $1/\e$ pole information as part of the simplification process,
but the simplest forms of the coefficients do not satisfy it manifestly.
That is, we shall obtain alternate, compact representations for the tree
amplitudes, by re-imposing the constraints from \eqn{OneLoopIRPoles}
after completing the simplifications.

It is also possible to relate box coefficients from one 
of the four seven-point NMHV helicity amplitudes to those from the other
three, using properties of the tree amplitudes which enter the cuts.
Such relations are possible any time a cut that is sensitive to a given
box integral does {\it not} contain the `nonsinglet-non-MHV' type of
contribution.   Under such circumstances, there are supersymmetry Ward
identities (SW)~\cite{SWI} which can be used to permute around the helicities
on one side of the cut.  

Consider for example the $s_{123}$ cut
of $A_{7;1}^{\NeqFour}(1^-,2^-,3^+,4^-,5^+,6^+,7^+)$.
It has a singlet-non-MHV term, which is given by
\begin{equation}
 \hbox{S-NMHV}_{123} = i 
   \, A^\tree_5((-\ell_1)^+,1^-,2^-,3^+,\ell_4^+)
   \, A^{\tree}_6((-\ell_4)^-,4^-,5^+,6^+,7^+,\ell_1^-) \,,
\label{SNMHV123Cut1}
\end{equation}
integrated over phase space.  Using a SWI on the left-side of the cut,
the cut becomes
\begin{equation}
 \hbox{S-NMHV}_{123} = i 
  { {\spa1.2}^4 \over {\spa2.3}^4 } 
  \, A^\tree_5((-\ell_1)^+,1^+,2^-,3^-,\ell_4^+)
   \, A^{\tree}_6((-\ell_4)^-,4^-,5^+,6^+,7^+,\ell_1^-)\,.
\label{SNMHV123Cut2}
\end{equation}
Except for the spinor-product prefactor, this expression is equivalent 
to the singlet-non-MHV piece of the $s_{234}$ cut
of $A_{7;1}^{\NeqFour}(1^-,2^-,3^-,4^+,5^+,6^+,7^+)$,
after the relabelling $1 \lr 4$, $2 \lr 3$, $5 \lr 7$ (and an overall
minus sign).
On the other hand, the nonsinglet-MHV term in the above $s_{123}$ cut
is given by 
\begin{eqnarray}
 \hbox{NS-MHV}_{123} &=& i 
  \sum_\lambda 
      A^\tree_5((-\ell_1)^{-\lambda},1^-,2^-,3^+,\ell_4^\lambda)
   \, A^{\tree}_6((-\ell_4)^{-\lambda},4^-,5^+,6^+,7^+,\ell_1^\lambda)
\nonumber \\
 &=& i { {\spba3.{(1+2)}.4}^4 \over  {\spba1.{(2+3)}.4}^4 }
    \sum_\lambda 
      A^\tree_5((-\ell_1)^{-\lambda},1^+,2^-,3^-,\ell_4^\lambda)
\nonumber \\
&& \hskip3.5cm
 \times A^{\tree}_6((-\ell_4)^{-\lambda},4^-,5^+,6^+,7^+,\ell_1^\lambda) \,.
\label{NSMHV123Cut2}
\end{eqnarray}
Again, except for the spinor-product prefactor, this expression is equivalent 
to the nonsinglet-MHV piece of the $s_{234}$ cut
of $A_{7;1}^{\NeqFour}(1^-,2^-,3^-,4^+,5^+,6^+,7^+)$,
after applying the same relabelling and minus sign.


\section{Results}
\label{ResultsSection}

In this section we present the results for the four independent
one-loop 7-point NMHV amplitudes in $\Neqfour$ SYM: 
$A_{7;1}^{\Neqfour}(1^-,2^-,3^-,4^+,5^+,6^+,7^+)$,
$A_{7;1}^{\Neqfour}(1^-,2^-,3^+,4^-,5^+,6^+,7^+)$,
$A_{7;1}^{\Neqfour}(1^-,2^-,3^+,4^+,5^-,6^+,7^+)$,
and 
$A_{7;1}^{\Neqfour}(1^-,2^+,3^-,4^+,5^-,6^+,7^+)$.
All other $A_{7;1}^{\Neqfour}$ NMHV helicity amplitudes can 
be obtained from these by the three operations:
\begin{enumerate}
\item {\it parity}, which exchanges $+$ and $-$ helicities
and is implemented on the basic spinor products by the action
$\spa{i}.{j} \lr \spb{j}.{i}$,
\item {\it reflection symmetry}, essentially charge conjugation invariance,
which states that 
$A_{n;1}^{\Neqfour}(1^{\lambda_1},2^{\lambda_2},\ldots,n^{\lambda_n})
= (-1)^n A_{n;1}^{\Neqfour}(n^{\lambda_n},\ldots,2^{\lambda_2},1^{\lambda_1})$,
and
\item {\it cyclic symmetry}, since each $A_{n;1}^{\Neqfour}$ is the coefficient
of a cyclicly-invariant color trace.
\end{enumerate}
The subleading-color partial amplitudes appearing in \eqn{ColorDecomposition}, 
$A_{7;c}^{\Neqfour}$ for $c>1$,
are obtained by sums over permutations of the leading-color
$A_{7;1}^{\Neqfour}$~\cite{Neq4Oneloop}, using \eqn{SublAnswer}.

\begin{figure}[t]
\centerline{\epsfxsize 6.0 truein \epsfbox{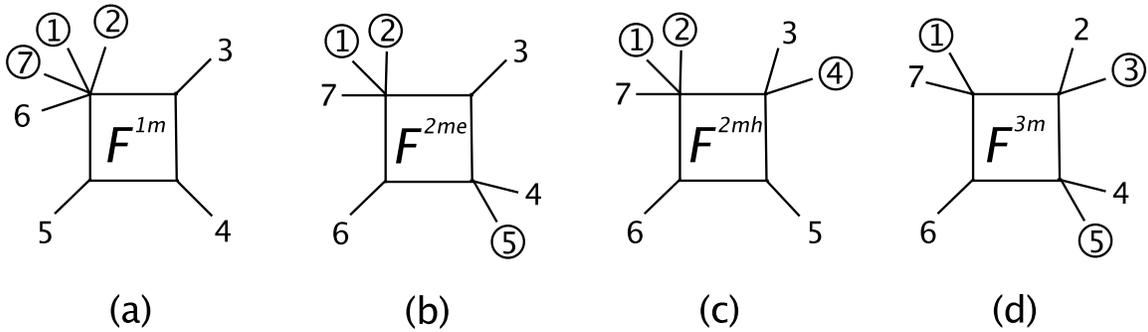}}
\caption[a]{\small Examples of box integral functions $B(i,j,k)$
appearing in 7-point amplitudes; the arguments $i,j,k$ are circled: 
(a) the one-mass box $B(1,2,7)=F^{\rm 1m}(s_{34},s_{45},s_{345})$,
(b) the `easy' two-mass box 
$B(1,2,5) = F^{{\rm 2m}e}(s_{345},s_{456},s_{45},s_{712})$,
(c) the `hard' two-mass box
$B(1,2,4) = F^{{\rm 2m}h}(s_{56},s_{345},s_{712},s_{34})$,
and (d) the three-mass box
$B(1,3,5) = F^{{\rm 3m}}(s_{671},s_{456},s_{71},s_{23},s_{45})$.}
\label{boxexamplesFigure}
\end{figure}

The $\Neqfour$ SYM amplitude is expressible as a sum of scalar box
integrals ${\cal I}_4$, or equivalently the box functions $F$
given in \app{IntegralsAppendix}.
For the color-ordered amplitudes $A_{7;1}$, the kinematics of 
each box integral can be obtained from the heptagon diagram
with external legs in the order 1,2,3,4,5,6,7, by deleting
three different propagators, say $i,j,k$.   We label each box 
function $F$, and each kinematic coefficient $c$ multiplying it,
by this triplet of integers.   However, to avoid confusion with 
a twistor-space `line operator' $F_{ijk}$ to appear in
\sect{TwistorSection}, we call these box functions $B(i,j,k)$
instead of $F(i,j,k)$.
Thus we write the leading-color partial amplitude as
\be
A_{7;1}^{\Neqfour} = i \cg \, (\mu^2)^\e
\sum_{i,j,k} c_{ijk} B(i,j,k)\,,
\label{GenBoxDecomp}
\ee
where
\be
c_\Gamma\ =\ {1 \over (4 \pi)^{2-\e}}
{\Gamma(1+\e)\Gamma^2(1-\e)\over\Gamma(1-2\e)}
\label{cGamma}
\ee
is a ubiquitous prefactor.
We label the propagators so that in the heptagon diagram the 
external leg labeled $p$ lies between propagators labeled 
$p$ and $p+1$~(mod 7).
Thus in the box $B(i,j,k)$, leg $i$ is grouped into a single massive 
leg with leg $i-1$, and similarly for legs
$j$ and $k$.  If two members of the set $\{i,j,k\}$ are cyclicly
adjacent, the corresponding massive leg of the box integral contains 
three massless legs.  If all three of $\{i,j,k\}$ are cyclicly adjacent, 
that massive leg contains four massless legs.
Some examples of this labeling are shown in \fig{boxexamplesFigure}.

The kinematic arguments of the boxes are $s$ and $t$, the squares
of the sums of momenta emerging from two adjacent vertices of the box,
followed by the `masses', the squares of the sums of momenta emerging 
from individual vertices of the box (when these are nonzero).
The easy two-mass box has two reflection symmetries, so it doesn't
matter which invariant is $s$ and which $t$.  For the hard two-mass box,
$s$ is the invariant formed by the two adjacent massless legs.
For the three-mass box, $s$ is the invariant formed by the massless leg
and the first massive leg following it cyclicly.

For a given helicity amplitude, the number of box functions, and box
coefficients, is the number of un-ordered integer triplets $(i,j,k)$, where 
each integer runs from 1 to 7, and all three are unequal.
This number is just $( {7 \atop 3} )$, or 35.
Another way to count this number is to observe that there are 5 different
kinds of box in the 7-point case:
\begin{itemize}
\item the one-mass box shown in \fig{boxexamplesFigure}a, plus
cyclic permutations (7 boxes),
\item the easy-two-mass box shown in \fig{boxexamplesFigure}b, plus
cyclic permutations (7 boxes),
\item the hard-two-mass box shown in \fig{boxexamplesFigure}c, plus its
`reflection' in which the 2-leg cluster precedes the 3-leg cluster, 
plus cyclic permutations (14 boxes),
\item the three-mass box shown in \fig{boxexamplesFigure}d, plus
cyclic permutations (7 boxes).
\end{itemize}

In the remainder of this section we present the coefficients $c_{ijk}$
of all the box functions, using symmetries whenever possible.
In section~\ref{TwistorSection} we shall describe the twistor-space 
structure and other general properties of these coefficients, 
which fit nicely into a uniform pattern. 


\subsection{$A_{7;1}^{\Neqfour}(1^-,2^-,3^-,4^+,5^+,6^+,7^+)$}
\label{mmmppppSection}

This partial amplitude possesses a reflection, or ``flip'' 
symmetry.  We define the flip operation (for this partial amplitude)
by
\be
X |_{\rm flip} = - X(1 \lr 3, 4 \lr 7, 5 \lr 6)\,,
\label{mmmppppflip}
\ee
{\it i.e.} a two-fold exchange of labels accompanied by an 
overall minus sign.
The tree amplitude obeys
\be
  A_{7}^{\tree}(1^-,2^-,3^-,4^+,5^+,6^+,7^+) \Bigr|_{\rm flip} 
= A_{7}^{\tree}(1^-,2^-,3^-,4^+,5^+,6^+,7^+)\,,
\label{mmmpppptreeflip}
\ee
as does the one-loop amplitude,
\be
  A_{7;1}^{\Neqfour}(1^-,2^-,3^-,4^+,5^+,6^+,7^+) \Bigr|_{\rm flip} 
= A_{7;1}^{\Neqfour}(1^-,2^-,3^-,4^+,5^+,6^+,7^+)\,.
\label{mmmppppampflip}
\ee

We can write all of the box coefficients in terms of the following quantities:
\bea
c_A &=& 
{ - {\spa2.3}^3 {\spba5.{(6+7)}.1}^3 \over 
  s_{234} s_{671} \spa3.4 \spa6.7 \spa7.1 
   \spba5.{(3+4)}.2 \spaa6.{(7+1)}.{(2+3)}.4 }  \ ,
\label{mmmppppcA} \\
c_B &=&
 { - {\spaa3.{(4+5)}.{(6+7)}.1}^3 
  \over s_{345} s_{671} \spa3.4 \spa4.5 \spa6.7 \spa7.1
    \spba2.{(3+4)}.5 \spba2.{(7+1)}.6 } \ ,
\label{mmmppppcB} \\
c_{136} &=& 
{ {\spa2.3}^3 {\spba7.{(5+6)}.4}^3
  \over  \spa3.4 \spa4.5 \spa5.6 \spb7.1 \spba1.{(2+3)}.4 
  \spaa6.{(7+1)}.{(2+3)}.4 \spaa2.{(7+1)}.{(5+6)}.4 } ,
\nonumber \\
\label{mmmppppc136} \\
c_{146} &=&
   {\spb5.6}^3 \times { {\spa1.2}^3 
 \over \spa3.4 \spba5.{(3+4)}.2 \spaa2.{(3+4)}.{(5+6)}.7 }
\nonumber \\
&& \hskip1.3cm \times
{ {\spa2.3}^3 
 \over \spa7.1 \spba6.{(7+1)}.2 \spaa2.{(7+1)}.{(5+6)}.4 } \ ,
\label{mmmppppc146} \\ 
c_{147} &=&
   { {\spa2.3}^3 {\spb6.7}^3 
 \over s_{671} \spa3.4 \spa4.5 \spb7.1 \spba1.{(6+7)}.5 \spba6.{(7+1)}.2 } \ ,
\label{mmmppppc147} \\ 
c_{247} &=&
- { {\spaa5.{(6+7)}.{(1+2)}.3}^3
 \over \spb1.2 \spa3.4 \spa4.5 \spa5.6 \spa6.7
   \spba1.{(6+7)}.5 \spba2.{(3+4)}.5 \spaa7.{(1+2)}.{(3+4)}.5 } \ ,
\nonumber \\
\label{mmmppppc247} \\ 
c_{346} &=&
- { {\spa2.3}^3 s_{567}^3 
  \over s_{234} \spa3.4 \spa5.6 \spa6.7 
  \spba1.{(2+3)}.4 \spba1.{(6+7)}.5 \spaa2.{(3+4)}.{(5+6)}.7 } \ ,
\label{mmmppppc346} \\ 
c_{347} &=&
   { {\spba4.{(2+3)}.1}^3
  \over s_{234} \spb2.3 \spb3.4 \spa5.6 \spa6.7 \spa7.1 
   \spba2.{(3+4)}.5 } \ ,
\label{mmmppppc347} \\
c_{567} &=&
   { s_{123}^3 
  \over \spb1.2 \spb2.3 \spa4.5 \spa5.6 \spa6.7
   \spba1.{(2+3)}.4 \spba3.{(1+2)}.7 } \ .
\label{mmmppppc567}
\eea
Note that $c_B$ and $c_{567}$ are flip symmetric.

The quantities~(\ref{mmmppppcA})--(\ref{mmmppppc567}), 
plus their images under the flip operation, are not all independent.  
They satisfy
\bea
[ c_{136} ] |_{\rm flip} + c_{346} + c_{567} &=& c_{247} + c_{347} \,,
\label{mmmppppEq3} \\
c_{146} + c_{147} + c_{346} &=& c_A + c_{136} \,,
\label{mmmppppEq4}
\eea
and
\bea
A_{7}^{\tree} &=&
[ c_A + c_{136} ] |_{\rm flip} + c_{147} + c_{346} + c_{567} 
\label{mmmpppptreeEq1} \\
&=&
 c_B + c_{347} + [ c_{347} ] |_{\rm flip} \,.
\label{mmmpppptreeEq2}
\eea
\Eqn{mmmpppptreeEq2} in particular provides a compact representation of the tree
amplitude which has the proper collinear behavior manifest in all
channels, resembling some ``twistor-space'' constructions~\cite{CSW}.
\Eqn{mmmpppptreeEq2} is manifestly flip symmetric, whereas
\eqn{mmmpppptreeEq1} is not.

In terms of these quantities, the remaining box coefficients are given by
\bea
c_{235} &=& c_{237} = c_{256} = c_{257} = c_{357} = c_{367} = 0,
\label{mmmppppczeroes} \\ 
c_{236} &=& c_{267} = c_{356} = c_{567},
\label{mmmppppc567s} \\ 
c_{134} &=&
 c_A + c_{147},
\label{mmmppppc134} \\ 
c_{137} &=&
 c_A + c_{347},
\label{mmmppppc137} \\ 
c_{167} &=&
 c_A + c_{136} + c_{347},
\label{mmmppppc167} \\ 
c_{234} &=&
 c_{247} + c_{347},
\label{mmmppppc234} \\
c_{345} &=&
 c_B + c_{147} + c_{347},
\label{mmmppppc345} \\
c_{457} &=&
 c_B + c_{147},
\label{mmmppppc457} \\
c_{467} &=&
 c_{346} + c_{347},
\label{mmmppppc467}
\eea 
and
\bea
c_{123} &=&
 c_{234} |_{\rm flip} \,,
\label{mmmppppc123} \\ 
c_{124} &=&
 c_{134} |_{\rm flip} \,,
\label{mmmppppc124} \\ 
c_{125} &=&
 c_{347} |_{\rm flip} \,,
\label{mmmppppc125} \\ 
c_{126} &=&
 c_{346} |_{\rm flip} \,,
\label{mmmppppc126} \\ 
c_{127} &=&
 c_{345} |_{\rm flip} \,,
\label{mmmppppc127} \\
c_{135} &=&
 c_{247} |_{\rm flip} \,,
\label{mmmppppc135} \\ 
c_{145} &=&
 c_{147} |_{\rm flip} \,,
\label{mmmppppc145} \\ 
c_{156} &=&
 c_{467} |_{\rm flip} \,,
\label{mmmppppc156} \\ 
c_{157} &=&
 c_{457} |_{\rm flip} \,,
\label{mmmppppc157} \\ 
c_{245} &=&
 c_{137} |_{\rm flip} \,,
\label{mmmppppc245} \\ 
c_{246} &=&
 c_{136} |_{\rm flip} \,,
\label{mmmppppc246} \\ 
c_{456} &=&
 c_{167} |_{\rm flip} \,.
\label{mmmppppc456}
\eea
The flip relations can also be summarized as
\be
c_{ijk} = c_{\tilde{\imath}\tilde{\jmath}\tilde{k}} |_{\rm flip} \,,
\label{gencflip} \\ 
\ee
where
\be
\tilde{\imath} = (5 - i) \hbox{ mod 7},
\label{mmmppppiflip}
\ee
and $\tilde{\imath}\tilde{\jmath}\tilde{k}$ should be written in ascending order.


\subsection{$A_{7;1}^{\Neqfour}(1^-,2^-,3^+,4^-,5^+,6^+,7^+)$}
\label{mmpmpppSection}

This partial amplitude is alone among the NMHV seven-point amplitudes
in having no flip symmetry.  The box coefficients $c_{ijk}$
are given in terms of the following quantities:
\bea
c_A &=&
- { {\spa2.4}^4 {\spba5.{(6+7)}.1}^3 
  \over s_{671} s_{234}
   \spa2.3 \spa3.4 \spa6.7 \spa7.1
   \spba5.{(3+4)}.2 \spaa6.{(7+1)}.{(2+3)}.4 } \ ,
\label{mmpmpppcA} \\
c_B &=&
   { {\spa1.2}^3 {\spba7.{(5+6)}.4}^3
  \over s_{123} s_{456}
  \spa2.3 \spa4.5 \spa5.6
  \spba7.{(1+2)}.3 \spaa6.{(4+5)}.{(2+3)}.1 } \ ,
\label{mmpmpppcB} \\
c_C &=&
{ - {\spaa4.{(3+5)}.{(6+7)}.1}^4 
  \over s_{345} s_{671}
  \spa3.4 \spa4.5 \spa6.7 \spa7.1
   \spba2.{(3+4)}.5 \spba2.{(7+1)}.6 \spaa3.{(4+5)}.{(6+7)}.1 } ,
\nonumber\\
\label{mmpmpppcC} \\
c_D &=&
   { {\spa1.2}^3 {\spba6.{(3+5)}.4}^4 
  \over s_{345} s_{712}
  \spa3.4 \spa4.5 \spa7.1
  \spba6.{(4+5)}.3 \spba6.{(7+1)}.2 \spaa5.{(3+4)}.{(1+2)}.7 } \ ,
\nonumber\\
\label{mmpmpppcD} \\
c_E &=&
   { {\spa1.2}^3 {\spba3.{(5+6)}.4}^4 
  \over s_{456} s_{712}
  \spa4.5 \spa5.6 \spa7.1 
  \spba3.{(4+5)}.6 \spba3.{(1+2)}.7 \spaa2.{(7+1)}.{(5+6)}.4 } \ ,
\nonumber\\
\label{mmpmpppcE} \\
c_{125} &=&
   { {\spba7.{(1+2)}.4}^4
  \over s_{712}
  \spb1.2 \spa3.4 \spa4.5 \spa5.6 \spb7.1
  \spba2.{(7+1)}.6 \spba7.{(1+2)}.3 } \ ,
\label{mmpmpppc125} \\
c_{135} &=& 
  { \bigl( \spba3.{(6+7)}.1 \spa4.6 + \spba3.{5}.4 \spa1.6 \bigr)^4
  \over \spb2.3 \spa4.5 \spa5.6 \spa6.7 \spa7.1
  \spba3.{(4+5)}.6 \spba2.{(7+1)}.6 }
\nonumber \\
&&\hskip0.5cm 
\times 
 { 1 \over \spaa6.{(7+1)}.{(2+3)}.4 \spaa6.{(4+5)}.{(2+3)}.1 } \ ,
\label{mmpmpppc135} \\
c_{136} &=&
   { {\spa2.4}^4 {\spba7.{(5+6)}.4}^3
 \over \spa2.3 \spa3.4 \spa4.5 \spa5.6 \spb7.1
 \spba1.{(2+3)}.4 }
\nonumber \\
&&\hskip0.5cm 
\times 
 { 1 \over \spaa2.{(7+1)}.{(5+6)}.4 \spaa6.{(7+1)}.{(2+3)}.4 } \ ,
\label{mmpmpppc136} \\
c_{145} &=&
   { {\spa1.2}^3 {\spb3.5}^4
 \over s_{345}
 \spb3.4 \spb4.5 \spa6.7 \spa7.1
 \spba3.{(4+5)}.6 \spba5.{(3+4)}.2 } \ ,
\label{mmpmpppc145} \\
c_{146} &=&
   { {\spa1.2}^3 {\spa2.4}^4 {\spb5.6}^3
 \over \spa2.3 \spa3.4 \spa7.1
 \spba5.{(3+4)}.2 \spba6.{(7+1)}.2 }
\nonumber \\
&&\hskip0.5cm 
\times 
 { 1 \over \spaa2.{(3+4)}.{(5+6)}.7 \spaa2.{(7+1)}.{(5+6)}.4 } \ ,
\label{mmpmpppc146} \\
c_{147} &=&
   { {\spa2.4}^4 {\spb6.7}^3 
 \over s_{671}
 \spa2.3 \spa3.4 \spa4.5 \spb7.1
  \spba1.{(6+7)}.5 \spba6.{(7+1)}.2 } \ ,
\label{mmpmpppc147} \\
c_{236} &=&
   { {\spba3.{(1+2)}.4}^4
 \over s_{123}
 \spb1.2 \spb2.3 \spa4.5 \spa5.6 \spa6.7
 \spba1.{(2+3)}.4 \spba3.{(1+2)}.7 } \ ,
\label{mmpmpppc236} \\
c_{237} &=&
   { {\spa1.2}^3 s_{567}^3
 \over s_{123}
 \spa2.3 \spa5.6 \spa6.7
 \spba4.{(2+3)}.1 \spba4.{(5+6)}.7 \spaa3.{(1+2)}.{(6+7)}.5 } \ ,
\label{mmpmpppc237} \\
c_{246} &=&
   { {\spa1.2}^3 {\spba3.{(5+6)}.7}^4
 \over \spb3.4 \spa5.6 \spa6.7 \spa7.1 
 \spba3.{(1+2)}.7 \spba4.{(5+6)}.7 }
\nonumber \\
&&\hskip0.5cm 
\times 
 { 1 \over \spaa2.{(3+4)}.{(5+6)}.7 \spaa5.{(3+4)}.{(1+2)}.7 } \ ,
\label{mmpmpppc246} \\
c_{247} &=&
   { {\spaa4.{(1+2)}.{(6+7)}.5}^4
 \over \spb1.2 \spa3.4 \spa4.5 \spa5.6 \spa6.7
 \spba1.{(6+7)}.5 \spba2.{(3+4)}.5 }
\nonumber \\
&&\hskip0.5cm 
\times 
 { 1 \over \spaa7.{(1+2)}.{(3+4)}.5 \spaa3.{(1+2)}.{(6+7)}.5 } \ ,
\label{mmpmpppc247} \\
c_{256} &=&
   { {\spa1.2}^3 {\spb5.6}^3 
 \over s_{456}
 \spa2.3 \spb4.5 \spa7.1
 \spba4.{(5+6)}.7 \spba6.{(4+5)}.3 } \ ,
\label{mmpmpppc256} \\
c_{257} &=&
   { {\spa1.2}^3 {\spa3.4}^3 {\spb6.7}^3
 \over \spa2.3 \spa4.5
 \spba7.{(1+2)}.3 \spba6.{(4+5)}.3 }
\nonumber \\
&&\hskip0.5cm 
\times 
 { 1 \over \spaa5.{(6+7)}.{(1+2)}.3 \spaa1.{(6+7)}.{(4+5)}.3 } \ ,
\label{mmpmpppc257} \\
c_{346} &=&
- { {\spa2.4}^4 s_{567}^3 
 \over s_{234}
 \spa2.3 \spa3.4 \spa5.6 \spa6.7
 \spba1.{(2+3)}.4 \spba1.{(6+7)}.5 \spaa2.{(3+4)}.{(5+6)}.7 } \ ,
\nonumber\\
\label{mmpmpppc346} \\
c_{347} &=&
   { {\spba3.{(2+4)}.1}^4 
 \over s_{234}
 \spb2.3 \spb3.4 \spa5.6 \spa6.7 \spa7.1
 \spba2.{(3+4)}.5 \spba4.{(2+3)}.1 } \ ,
\label{mmpmpppc347} \\
c_{357} &=&
   { {\spa1.2}^3 {\spba5.{(6+7)}.1}^3 
 \over \spa2.3 \spb4.5 \spa6.7 \spa7.1
 \spba4.{(2+3)}.1 }
\nonumber \\
&&\hskip0.5cm 
\times 
 { 1 \over \spaa3.{(4+5)}.{(6+7)}.1 \spaa6.{(4+5)}.{(2+3)}.1 } \ .
\label{mmpmpppc357}
\eea

Again the quantities~(\ref{mmpmpppcA})--(\ref{mmpmpppc357}) are not 
independent.  They satisfy
\bea
c_A + c_{136} &=& c_{146} + c_{147} + c_{346},
\label{mmpmpppEq1} \\
c_C + c_{125} + c_{237} + c_{257} + c_{347}
 &=& c_D + c_{147} + c_{236} + c_{246} + c_{346},
\label{mmpmpppEq2prime} \\
c_D + c_{246} + c_{256} &=& c_E + c_{145} + c_{146},
\label{mmpmpppEq3} \\
c_{236} + c_{246} + c_{346} &=& c_{237} + c_{247} + c_{347},
\label{mmpmpppEq4} \\
c_B + c_{357} &=& c_{237} + c_{256} + c_{257},
\label{mmpmpppEq5} \\
c_A + c_{135} + c_{145} &=& c_C + c_{347} + c_{357}.
\label{mmpmpppEq6}
\eea
The tree amplitude may be written in terms of the box coefficients
in several different ways.  The shortest expression we have found is:
\be
A_{7}^{\tree} =
 c_B + c_C + c_{125} + c_{347} + c_{357}.
\label{mmpmppptreeEq2}
\ee

The remaining box coefficients are given by
\bea
c_{123} &=&
 c_E + c_{136} + c_{236} + c_{256},
\label{mmpmpppc123} \\
c_{124} &=&
 c_D + c_{145},
\label{mmpmpppc124} \\
c_{126} &=&
 c_E + c_{256},
\label{mmpmpppc126} \\
c_{127} &=&
 c_C + c_{125} + c_{145} + c_{257},
\label{mmpmpppc127} \\
c_{134} &=&
 c_A + c_{147},
\label{mmpmpppc134} \\
c_{137} &=&
 c_A + c_{347},
\label{mmpmpppc137} \\
c_{156} &=&
 c_E + c_{125},
\label{mmpmpppc156} \\
c_{157} &=&
 c_C + c_{145},
\label{mmpmpppc157} \\
c_{167} &=&
 c_A + c_{136} + c_{347},
\label{mmpmpppc167} \\
c_{234} &=&
 c_{236} + c_{246} + c_{346},
\label{mmpmpppc234} \\
c_{235} &=&
 c_B + c_{256},
\label{mmpmpppc235} \\
c_{245} &=&
 c_D + c_{125},
\label{mmpmpppc245} \\
c_{267} &=&
 c_{236} + c_{237},
\label{mmpmpppc267} \\
c_{345} &=&
 c_C + c_{147} + c_{347} + c_{357},
\label{mmpmpppc345} \\
c_{356} &=&
 c_B + c_{236},
\label{mmpmpppc356} \\
c_{367} &=&
 0,
\label{mmpmpppc367} \\
c_{456} &=&
 c_D + c_{125} + c_{246} + c_{256},
\label{mmpmpppc456} \\
c_{457} &=&
 c_C + c_{147},
\label{mmpmpppc457} \\
c_{467} &=&
 c_{346} + c_{347},
\label{mmpmpppc467} \\
c_{567} &=&
 c_B + c_{236} + c_{357}.
\label{mmpmpppc567}
\eea


\subsection{$A_{7;1}^{\Neqfour}(1^-,2^-,3^+,4^+,5^-,6^+,7^+)$}
\label{mmppmppSection}

For this partial amplitude we define the ``flip'' operation,
\be
X |_{\rm flip} = - X(1 \lr 2, 3 \lr 7, 4 \lr 6),
\label{mmppmppflip}
\ee
which is a symmetry of the tree amplitude,
\be
  A_{7}^{\tree}(1^-,2^-,3^+,4^+,5^-,6^+,7^+) \Bigr|_{\rm flip} 
= A_{7}^{\tree}(1^-,2^-,3^+,4^+,5^-,6^+,7^+) \,,
\label{mmppmpptreeflip}
\ee
and of the one-loop amplitude.  The box coefficients $c_{ijk}$
are given in terms of the following quantities:
\bea
c_A &=&
   { {\spa1.2}^3 {\spba7.{(4+6)}.5}^4
 \over s_{123} s_{456} 
 \spa2.3 \spa4.5 \spa5.6
 \spba7.{(5+6)}.4 \spba7.{(1+2)}.3 \spaa6.{(4+5)}.{(2+3)}.1 } \ ,
\nonumber\\
\label{mmppmppcA} \\
c_B &=&
   { {\spa1.2}^3 {\spba4.{(6+7)}.5}^4
 \over s_{123} s_{567}
 \spa2.3 \spa5.6 \spa6.7
 \spba4.{(5+6)}.7 \spba4.{(2+3)}.1 \spaa3.{(1+2)}.{(6+7)}.5 } \ ,
\nonumber\\
\label{mmppmppcB} \\
c_C &=&
 - { {\spaa5.{(3+4)}.{(6+7)}.1}^4
 \over s_{345} s_{671}
 \spa3.4 \spa4.5 \spa6.7 \spa7.1
 \spba2.{(3+4)}.5 \spba2.{(7+1)}.6 }
\nonumber \\
&&\hskip0.5cm 
\times 
 { 1 \over \spaa3.{(4+5)}.{(6+7)}.1 } \,,
\nonumber\\
\label{mmppmppcC} \\
c_D &=&
   { {\spaa1.{(6+7)}.{(3+4)}.2}^4
 \over s_{234} s_{671}
 \spa2.3 \spa3.4 \spa6.7 \spa7.1
 \spba5.{(3+4)}.2 \spba5.{(6+7)}.1 }
\nonumber \\
&&\hskip0.5cm 
\times 
 { 1 \over \spaa4.{(2+3)}.{(7+1)}.6 } \ ,
\label{mmppmppcD} \\
c_{136} &=& 
  { \bigl( \spba7.{(3+4)}.2 \spa5.4 + \spba7.{6}.5 \spa2.4 \bigr)^4
 \over \spa2.3 \spa3.4 \spa4.5 \spa5.6 \spb7.1
 \spba1.{(2+3)}.4 \spba7.{(5+6)}.4 }
\nonumber \\
&&\hskip0.5cm 
\times 
 { 1 \over \spaa6.{(7+1)}.{(2+3)}.4 \spaa2.{(7+1)}.{(5+6)}.4 } \ ,
\label{mmppmppc136} \\
c_{147} &=&
   { {\spa2.5}^4 {\spb6.7}^3
 \over s_{671}
 \spa2.3 \spa3.4 \spa4.5 \spb7.1
 \spba1.{(6+7)}.5 \spba6.{(7+1)}.2 } \ ,
\label{mmppmppc147} \\
c_{236} &=&
   { {\spba3.{(1+2)}.5}^4
 \over s_{123}
 \spb1.2 \spb2.3 \spa4.5 \spa5.6 \spa6.7
 \spba3.{(1+2)}.7 \spba1.{(2+3)}.4 } \ ,
\label{mmppmppc236} \\
c_{247} &=&
   { {\spaa5.{(3+4)}.{(6+7)}.5}^4
 \over \spb1.2 \spa3.4 \spa4.5 \spa5.6 \spa6.7
 \spba2.{(3+4)}.5 \spba1.{(6+7)}.5 }
\nonumber \\
&&\hskip0.5cm 
\times 
 { 1 \over \spaa5.{(6+7)}.{(1+2)}.3 \spaa5.{(3+4)}.{(1+2)}.7 } \ ,
\label{mmppmppc247} \\
c_{256} &=&
   { {\spa1.2}^3 {\spb4.6}^4
 \over s_{456}
 \spa2.3 \spb4.5 \spb5.6 \spa7.1
 \spba4.{(5+6)}.7 \spba6.{(4+5)}.3 } \ ,
\label{mmppmppc256} \\
c_{257} &=&
   { {\spa1.2}^3 {\spa3.5}^4 {\spb6.7}^3
 \over \spa2.3 \spa3.4 \spa4.5
 \spba6.{(4+5)}.3 \spba7.{(1+2)}.3 }
\nonumber \\
&&\hskip0.5cm 
\times 
 { 1 \over \spaa3.{(4+5)}.{(6+7)}.1 \spaa3.{(1+2)}.{(6+7)}.5 } \ ,
\label{mmppmppc257} \\
c_{357} &=&
   { {\spa1.2}^3 {\spba4.{(6+7)}.1}^4
 \over \spa2.3 \spb4.5 \spa6.7 \spa7.1
 \spba4.{(2+3)}.1 \spba5.{(6+7)}.1 }
\nonumber \\
&&\hskip0.5cm 
\times 
 { 1 \over \spaa3.{(4+5)}.{(6+7)}.1 \spaa6.{(4+5)}.{(2+3)}.1 } \ ,
\label{mmppmppc357} \\
c_{367} &=&
   { {\spa1.2}^3 {\spb6.7}^3
 \over s_{567}
 \spa2.3 \spa3.4 \spb5.6 \spba5.{(6+7)}.1 \spba7.{(5+6)}.4 } \ .
\label{mmppmppc367}
\eea
Note that $c_D$ and $c_{256}$ are flip-symmetric.

The quantities~(\ref{mmppmppcA})--(\ref{mmppmppc367}) obey the linear 
relations (plus their flips),
\bea
c_A + c_{357} + c_{367} &=& c_B + c_{256} + c_{257},
\label{mmppmppEq1} \\
c_A + c_C - c_{147} - c_{236} + c_{357} + c_{367} 
 &=& [ c_A + c_C - c_{147} - c_{236} + c_{357} + c_{367} ] |_{\rm flip} \,,
\label{mmppmppEq3} \\
c_C + c_{257} &=& [ c_B - c_{236} ] |_{\rm flip} + c_{147} + c_{247},
\label{mmppmppEq4} \\
c_D + c_{136} + c_{367} &=& [ c_C + c_{357} ] |_{\rm flip} + c_{147},
\label{mmppmppEq5} \\
c_A - c_{136} - c_{236} &=& [ c_A - c_{136} - c_{236} ] |_{\rm flip} \,.
\label{mmppmppEq6}
\eea
The tree amplitude can be written in a manifestly flip-symmetric form as
\be 
A_7^\tree =
 c_{247} + c_{256}
+ c_B + c_{147} + [ c_B + c_{147} ] |_{\rm flip} \,.
\label{mmppmpptreeEq3}
\ee

The remaining independent box coefficients are given by
\bea
c_{123} &=&
 [ c_A ] |_{\rm flip} + c_{136} + c_{236} + c_{256},
\label{mmppmppc123} \\
c_{127} &=&
 c_C + c_{257} + [ c_{236} + c_{367} ] |_{\rm flip} \,,
\label{mmppmppc127} \\
c_{134} &=&
 c_D + c_{147},
\label{mmppmppc134} \\
c_{157} &=&
 c_C + [ c_{367} ] |_{\rm flip} \,,
\label{mmppmppc157} \\
c_{235} &=&
 c_A + c_{256},
\label{mmppmppc235} \\
c_{237} &=&
 c_B + c_{367},
\label{mmppmppc237} \\
c_{267} &=&
 c_B + c_{236},
\label{mmppmppc267} \\
c_{345} &=&
 c_C + c_{147} + [ c_{147} ] |_{\rm flip} + c_{357},
\label{mmppmppc345} \\
c_{356} &=&
 c_A + c_{236},
\label{mmppmppc356} \\
c_{457} &=&
 c_C + c_{147},
\label{mmppmppc457} \\
c_{567} &=&
 c_A + c_{236} + c_{357} + c_{367}.
\label{mmppmppc567}
\eea
The final 16 box coefficients are obtained by the flip symmetry
via \eqn{gencflip}, where
\be
\tilde{\imath} = (4 - i) \hbox{ mod 7},
\label{mmppmppiflip}
\ee
and again $\tilde{\imath}\tilde{\jmath}\tilde{k}$ should be written 
in ascending order.


\subsection{$A_{7;1}^{\Neqfour}(1^-,2^+,3^-,4^+,5^-,6^+,7^+)$}
\label{mpmpmppSection}

For this partial amplitude we define the ``flip'' operation,
\be
X |_{\rm flip} = - X(1 \lr 5, 2 \lr 4, 6 \lr 7),
\label{mpmpmppflip}
\ee
which is a symmetry of the tree amplitude,
\be
  A_{7}^{\tree}(1^-,2^+,3^-,4^+,5^-,6^+,7^+) \Bigr|_{\rm flip} 
= A_{7}^{\tree}(1^-,2^+,3^-,4^+,5^-,6^+,7^+) \,,
\label{mpmpmpptreeflip}
\ee
and of the one-loop amplitude.  The box coefficients $c_{ijk}$
are given in terms of the following quantities:
\bea
c_A &=&
   { {\spa1.3}^4 {\spba7.{(4+6)}.5}^4
 \over s_{123} s_{456}
 \spa1.2 \spa2.3 \spa4.5 \spa5.6
 \spba7.{(5+6)}.4 \spba7.{(1+2)}.3 \spaa6.{(4+5)}.{(2+3)}.1 } ,
\nonumber\\
\label{mpmpmppcA} \\
c_B &=&
   { {\spa1.3}^4 {\spba4.{(6+7)}.5}^4
 \over s_{123} s_{567}
 \spa1.2 \spa2.3 \spa5.6 \spa6.7
 \spba4.{(5+6)}.7 \spba4.{(2+3)}.1 \spaa3.{(1+2)}.{(6+7)}.5 } ,
\nonumber\\
\label{mpmpmppcB} \\
c_C &=&
   { {\spaa1.{(6+7)}.{(2+4)}.3}^4 
 \over s_{234} s_{671}
 \spa2.3 \spa3.4 \spa6.7 \spa7.1
 \spba5.{(3+4)}.2 \spba5.{(6+7)}.1 \spaa4.{(2+3)}.{(7+1)}.6 } ,
\nonumber\\
\label{mpmpmppcC} \\
c_D &=&
 { - {\spaa5.{(4+6)}.{(7+2)}.1}^4
 \over s_{456} s_{712}
 \spa1.2 \spa4.5 \spa5.6 \spa7.1
 \spba3.{(4+5)}.6 \spba3.{(1+2)}.7 \spaa4.{(5+6)}.{(7+1)}.2 } ,
\nonumber\\
\label{mpmpmppcD} \\
c_{136} &=& 
  { \bigl( \spba7.{(2+4)}.3 \spa5.4 + \spba7.{6}.5 \spa3.4 \bigr)^4
 \over \spa2.3 \spa3.4 \spa4.5 \spa5.6 \spb7.1
 \spba1.{(2+3)}.4 \spba7.{(5+6)}.4 }
\nonumber \\
&&\hskip0.5cm 
\times 
 { 1 \over \spaa4.{(5+6)}.{(7+1)}.2 \spaa4.{(2+3)}.{(7+1)}.6 } \ ,
\label{mpmpmppc136} \\
c_{236} &=&
   { {\spba2.{(1+3)}.5}^4
 \over s_{123}
 \spb1.2 \spb2.3 \spa4.5 \spa5.6 \spa6.7
 \spba3.{(1+2)}.7 \spba1.{(2+3)}.4 } \ ,
\label{mpmpmppc236} \\
c_{246} &=& 
  { \bigl( \spba4.{(2+7)}.1 \spa5.7 + \spba4.{6}.5 \spa1.7 \bigr)^4
 \over \spa1.2 \spb3.4 \spa5.6 \spa6.7 \spa7.1
 \spba4.{(5+6)}.7 \spba3.{(1+2)}.7 }
\nonumber \\
&&\hskip0.5cm 
\times 
 { 1 \over \spaa7.{(1+2)}.{(3+4)}.5 \spaa7.{(5+6)}.{(3+4)}.2 } \ ,
\label{mpmpmppc246} \\
c_{256} &=&
   { {\spa1.3}^4 {\spb4.6}^4
 \over s_{456}
 \spa1.2 \spa2.3 \spb4.5 \spb5.6 \spa7.1
 \spba4.{(5+6)}.7 \spba6.{(4+5)}.3 } \ ,
\label{mpmpmppc256} \\
c_{257} &=&
   { {\spa1.3}^4 {\spa3.5}^4 {\spb6.7}^3 
 \over \spa1.2 \spa2.3 \spa3.4 \spa4.5
 \spba6.{(4+5)}.3 \spba7.{(1+2)}.3 }
\nonumber \\
&&\hskip0.5cm 
\times 
 { 1 \over \spaa3.{(4+5)}.{(6+7)}.1 \spaa3.{(1+2)}.{(6+7)}.5 } \ ,
\label{mpmpmppc257} \\
c_{347} &=&
   { {\spb2.4}^4 {\spa1.5}^4
 \over s_{234}
 \spb2.3 \spb3.4 \spa5.6 \spa6.7 \spa7.1
 \spba2.{(3+4)}.5 \spba4.{(2+3)}.1 } \ ,
\label{mpmpmppc347} \\
c_{357} &=&
   { {\spa1.3}^4 {\spba4.{(6+7)}.1}^4
 \over \spa1.2 \spa2.3 \spb4.5 \spa6.7 \spa7.1
 \spba4.{(2+3)}.1 \spba5.{(6+7)}.1 }
\nonumber \\
&&\hskip0.5cm 
\times 
 { 1 \over \spaa3.{(4+5)}.{(6+7)}.1 \spaa6.{(4+5)}.{(2+3)}.1 } \ ,
\label{mpmpmppc357} \\
c_{367} &=&
   { {\spa1.3}^4 {\spb6.7}^3
 \over s_{567}
 \spa1.2 \spa2.3 \spa3.4 \spb5.6
 \spba5.{(6+7)}.1 \spba7.{(5+6)}.4 } \ .
\label{mpmpmppc367}
\eea
Note that $c_D$, $c_{257}$ and $c_{347}$ are flip symmetric.

The linear relations obeyed by the 
quantities~(\ref{mpmpmppcA})--(\ref{mpmpmppc367}) are
\bea
c_A + c_{357} + c_{367} &=& c_B + c_{256} + c_{257},
\label{mpmpmppEq1} \\
c_C + c_D + c_{136} + c_{367}
 &=& c_B - c_{236} + c_{256} + [ c_B - c_{236} + c_{256} ] |_{\rm flip}
+ c_{257} + c_{347},
\label{mpmpmppEq3} \\
c_C + c_{136} + c_{367} &=& [ c_C + c_{136} + c_{367} ] |_{\rm flip} \,,
\label{mpmpmppEq4} \\
c_B - c_{236} - c_{246} + c_{347} &=& [ c_C - c_{357} ] |_{\rm flip} \,,
\label{mpmpmppEq5}
\eea
plus the equations related by the flip symmetry.
The simplest, manifestly flip-symmetric representation
we have been able to find for the tree amplitude is
\be
A_7^\tree =
 c_{257} + c_{347}
+ c_B + c_{256} + [ c_B + c_{256} ] |_{\rm flip} \,.
\label{mpmpmpptreeEq5}
\ee

The remaining independent box coefficients are given by
\bea
c_{123} &=&
 c_D + c_{136} + c_{236} + c_{256},
\label{mpmpmppc123} \\
c_{126} &=&
 c_D + c_{256},
\label{mpmpmppc126} \\
c_{134} &=&
 c_C + [ c_{367} ] |_{\rm flip} \,,
\label{mpmpmppc134} \\
c_{137} &=&
 c_C + c_{347},
\label{mpmpmppc137} \\
c_{167} &=&
 c_C + c_{136} + c_{347} + c_{367},
\label{mpmpmppc167} \\
c_{234} &=&
 c_B + c_{347} + c_{367} + [ c_{357} ] |_{\rm flip} \,,
\label{mpmpmppc234} \\
c_{235} &=&
 c_A + c_{256},
\label{mpmpmppc235} \\
c_{237} &=&
 c_B + c_{367},
\label{mpmpmppc237} \\
c_{267} &=&
 c_B + c_{236},
\label{mpmpmppc267} \\
c_{356} &=&
 c_A + c_{236},
\label{mpmpmppc356} \\
c_{567} &=&
 c_A + c_{236} + c_{357} + c_{367}.
\label{mpmpmppc567}
\eea
The final 16 box coefficients are obtained by the flip
symmetry~(\ref{gencflip}), where
\be
\tilde{\imath} = (7 - i) \hbox{ mod 7},
\label{mpmpmppiflip}
\ee
and again $\tilde{\imath}\tilde{\jmath}\tilde{k}$ should be written 
in ascending order.


\section{Consistency of the Results}
\label{ConsistencySection}

In this section we describe various consistency checks that we
performed on the amplitudes.  Amplitudes, in general, must satisfy a
stringent set of constraints on their analytic properties.  In
particular, all kinematic poles in the amplitudes must correspond to the
physical propagation of particles.  Such constraints are so restrictive
that they can be used, for example, to construct ans\"atze for 
infinite sequences of all-plus-helicity amplitudes in gauge and gravity 
theories~\cite{AllPlus,AllPlusGrav}.  
The poles may be divided into three categories: collinear, multi-particle
and spurious.  We consider each of these categories in turn.

\subsection{Collinear Behavior}

An important constraint on the amplitudes arises from the region in
phase space where the momenta of two legs $a$ and $b$ become collinear. 
In the collinear region, $k_a \to z k_P$, $k_b \to (1-z) k_P$, 
where $k_P$ is the momentum of the quasi-on-shell intermediate state $P$,
with helicity $\lambda$.  In this limit, massless color-ordered tree 
amplitudes behave as
\begin{equation}
A_{n}^{\tree}\ \mathop{\longrightarrow}^{a \parallel b}\
\sum_{\lambda=\pm} 
\Split^\tree_{-\lambda}   (z, a^{\lambda_a},b^{\lambda_b})\,
         A_{n-1}^{\tree}(\ldots(a+b)^\lambda\ldots)\,,
\label{TreeSplit}
\end{equation}
where $\Split^\tree_{-\lambda}$ are tree-level splitting
amplitudes~\cite{TreeReview}. At one loop, the generalization is,
\begin{eqnarray}
&& A_{n}^{\oneloop}\ \mathop{\longrightarrow}^{a \parallel b}\
\sum_{\lambda=\pm}  \biggl(
\Split^\tree_{-\lambda}   (z, a^{\lambda_a},b^{\lambda_b})\,
         A_{n-1}^{\oneloop}(\ldots(a+b)^\lambda\ldots) \nonumber \\
& & \hskip 2.7 cm \null
 + \Split^{\oneloop}_{-\lambda}(z,a^{\lambda_a},b^{\lambda_b})\,
         A_{n-1}^\tree(\ldots(a+b)^\lambda\ldots) \biggr) \,,
\label{LoopSplit}
\end{eqnarray}
where the $\Split^{\oneloop}_{-\lambda}$ are one-loop splitting
amplitudes, which are tabulated in the second appendix of
ref.~\cite{Neq4Oneloop}.  This reference also contains a
discussion of the behavior of the collinear limits of one-loop
amplitudes and integral functions.

We have verified numerically that all helicity amplitudes presented in
\sect{ResultsSection} have the proper collinear behavior in all
channels, as dictated by \eqn{LoopSplit}.  This, by itself, provides a
rather powerful check on their form.


\subsection{Multi-Particle Factorization}

A related constraint is that of multi-particle factorization.
This factorization plays a central role in the tree-level
CSW construction.  Presumably, it will also
play an important role in extending the CSW construction to loop level.
These properties are not as well-discussed in the literature 
as the collinear properties, so we present the key points here,
and include an example in \app{FactorizationAppendix}.

At tree level, multi-particle factorization is reasonably straightforward.
For $(k_i + k_{i+1} + \cdots + k_{i+r-1})^2 \equiv K^2 \rightarrow 0$ 
(with $r>2$), the amplitude behaves as
\begin{equation}
A_{n}^{\tree}\
\mathop{\longrightarrow}^{K^2 \rightarrow 0}\
\sum_{\lambda=\pm}  
      A_{r+1}^{\tree}(k_i, \ldots, k_{i+r-1}, K^\lambda) \, {i \over K^2} \, 
      A_{n-r+1}^{\tree}((-K)^{-\lambda}, k_{i+r}, \ldots, k_{i-1})\,,
\nonumber
\end{equation}
where $\lambda$ is the helicity of the intermediate state with momentum
$K$.

\begin{figure}[t]
\centerline{\epsfxsize 4. truein \epsfbox{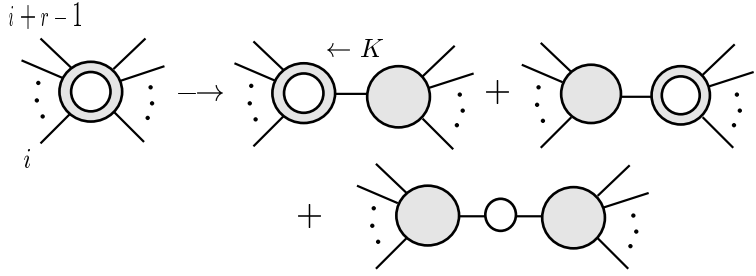}}
\caption{A schematic depiction of multi-particle 
factorization at one loop.}
\label{MultiFactFigure}
\end{figure}

At loop level, in infrared-divergent gauge theories, multi-particle
factorization is more subtle.  For such theories, loop amplitudes do
not factorize in any naive sense,  due to the emission of soft gluons
which induce non-trivial logarithmic corrections to 
factorization formul\ae.  These logarithmic corrections do however obey
universal formul\ae{}~\cite{BernChalmers} because of their close
relation to the universal infrared divergences~\cite{UniversalIR}.
More explicitly, for $K^2 \rightarrow 0$
the factorization properties for one-loop amplitudes are described 
by~\cite{BernChalmers},
\begin{eqnarray}
A_{n;1}^{\oneloop}\
&& \hskip -.42 cm 
 \mathop{\longrightarrow}^{K^2 \rightarrow 0} 
\hskip .15 cm 
\sum_{\lambda=\pm}  \Biggl[
      A_{r+1;1}^{\oneloop}(k_i, \ldots, k_{i+r-1}, K^\lambda) \, {i \over K^2} \, 
      A_{n-r+1}^{\tree}((-K)^{-\lambda}, k_{i+r}, \ldots, k_{i-1})
\nonumber \\
&& \hskip-.6cm \null
 +  A_{r+1}^{\tree}(k_i, \ldots, k_{i+r-1}, K^\lambda) \, {i\over K^2} \,
      A_{n-r+1;1}^{\oneloop}((-K)^{-\lambda}, k_{i+r}, \ldots, k_{i-1})
\label{LoopFact} \\
&& \hskip-.6cm \null 
 + A_{r+1}^{\tree}(k_i, \ldots, k_{i+r-1}, K^\lambda) \, {i\over K^2} \,
   A_{n-r+1}^{\tree}((-K)^{-\lambda}, k_{i+r}, \ldots, k_{i-1}) \, 
      \cg\,  \Fact_n(K^2;k_1, \ldots, k_n) \Biggr] \,,
\nonumber
\end{eqnarray}
where the one-loop {\it factorization function} $\Fact_n$ is
independent of helicities.  This formula is similar to the one for an
amplitude which factorizes naively, depicted in
\fig{MultiFactFigure}, except that $\Fact_n$ contains kinematic
invariants with momenta from both sides of the propagator carrying momentum 
$K$. For example $\ln(-s_{i-1, i})$ is one such
logarithm:  $k_{i-1}$ is a momentum belonging to one of the factorized
amplitudes on the right-hand side of \eqn{LoopFact} and $k_i$ to
the other amplitude. 

In general, the factorization function is composed of `factorizing'
and `non-factorizing' components,
\begin{equation}
\Fact_n = \Fact_n^{\rm fact} + \Fact_n^{\rm non\hbox{-}fact}\,.
\label{FactAndNonFact}
\end{equation}
The factorizing contributions are easily obtained by computing bubble
Feynman diagrams.  For the case of $\NeqFour$ supersymmetric gauge
theory the factorizing contributions vanish, leaving only the
`non-factorizing' contributions.

These `non-factorizing' contributions are linked to the infrared
divergences, as shown in ref.~\cite{BernChalmers}.  A constructive
proof of the universality of these non-factorizing contributions was
also given in that reference, as well as explicit formul\ae{} for
determining their values in any theory.  All non-factorizing
contributions are determined from the infrared divergences present in
a given process. The infrared divergences in the
$\NeqFour$ theory at one loop are
\begin{equation}
A_{n;1}^{\NeqFour}(1,2,\ldots,n)\Bigr|_{\e\ {\rm pole}}\ =\
 -{\cg\over\e^2} \sum_{j=1}^n
         \left( { \mu^2 \over -s_{j,j+1} } \right)^\e 
 A_{n}^{\tree}(1,2,\ldots,n) \ .
\label{NPtSingular}
\end{equation}
From Table 2 of ref.~\cite{BernChalmers}, the appearance of
the singularities, 
\begin{equation}
\Biggl[ -{\cg\over\e^2} \biggl( {\mu^2 \over -s_{i-1, i}} \biggr)^\e 
        -{\cg\over\e^2} \biggl( {\mu^2 \over -s_{i+r-1,i+r}} \biggr)^\e \Biggr]
 A_{n}^{\tree} \,,
\label{KeyMismatchIR}
\end{equation}
implies that in the factorization channel
$K^2 = s_{i \ldots (i+r-1)} \rightarrow 0$,
the factorization function is
\begin{eqnarray}
\Fact_n^{\Neqfour} (K^2; k_1, \ldots, k_n) &=&
 2 (\mu^2)^\eps \Bigl(  \Fhard{r-1;i+1} 
                      + \Fhard{n-r-1;i+r+1}
      + {2 \over \eps^2} (-s_{i\ldots (i+r-1)})^{-\eps} \Bigr) \nonumber \\
 &=& 2(\mu^2)^\eps\Bigl(F^{{\rm 2m}\, h}(s_{i-1,i}, s_{i\ldots (i+r-1)}, 
                           s_{(i+1)\ldots (i+r-1)},
                                         s_{(i+r)\ldots(i-2)} ) 
   \nonumber \\
&& \hskip 1.3 cm \null 
                    + F^{{\rm 2m}\, h} (s_{i+r-1, i+r},  s_{i \ldots (i+r-1)}, 
                                        s_{(i+r+1) \ldots (i-1)},
                                        s_{i \ldots (i+r-2)} ) \nonumber \\
&& \hskip 1.3 cm \null 
     + {2 \over \eps^2} (-s_{i \ldots (i+r-1)})^{-\eps} \Bigr)\,,  
 \label{GeneralFactorizationFunction}
\end{eqnarray}
where we use the notation from ref.~\cite{BernChalmers} on the first
line. On the subsequent lines we use the notation 
of \app{IntegralsAppendix} for the integral functions.

We have checked that all amplitudes in \sect{ResultsSection} satisfy
the correct multi-particle factorizations dictated by \eqn{LoopFact},
with the factorization function (\ref{GeneralFactorizationFunction}).
In \app{FactorizationAppendix}, we present an example
of a multi-particle factorization limit.  All the remaining 
multi-particle factorization limits, for any of the 
seven-point helicity amplitudes given in \sect{ResultsSection}, 
are similar.

\subsection{Cancellation of spurious singularities}

In addition to the collinear and multi-particle poles discussed in the
previous two subsections, the coefficients also contain `planar' and
`cubic' singularities as defined in \sect{CalculationSection}.  Unlike the
collinear and multi-particle singularities, these cannot be singularities
of the whole amplitude.  Factorization (even in massless gauge theories)
does not allow them.  We therefore expect them to cancel between different
integral functions.  Their origin in the integral reductions also suggests
this.  

This is indeed what one finds.  For example, in 
$A_{7;1}^{\Neqfour}(1^-,2^-,3^-,4^+,5^+,6^+,7^+)$, there are eight
coefficients that contain a denominator factor of $\spba1.{(2+3)}.4$.
As noted in \sect{CalculationSection}, these coefficients ---
$c_{136}$, 
$c_{167}$, 
$c_{236}$, 
$c_{267}$, 
$c_{346}$, 
$c_{356}$, 
$c_{467}$, and
$c_{567}$ --- will therefore be separately singular when 
$k_2+k_3 = a_1 k_1 + a_4 k_4$.  One can nonetheless verify
that these singularities cancel in the amplitude as a whole.  Indeed,
the cancellation occurs separately within two linear combinations of terms
in the amplitude,
\begin{eqnarray}
& & c_{136} B(1,3,6) + c_{236} B(2,3,6) + c_{346} B(3,4,6) 
  + c_{356} B(3,5,6) \,,
\nonumber\\
{{\rm and}\ \ \ \ } 
& & c_{167} B(1,6,7) + c_{267} B(2,6,7) + c_{467} B(4,6,7)
  + c_{567} B(5,6,7)\,.
\label{SingularityCancellation}
\end{eqnarray}
While a subset of terms in the whole amplitude will allow the cancellation
of any given singularity --- typically four terms for the planar 
singularities, and five for the cubic ones --- there does not appear to
be any subset of terms in which {\it all\/} spurious singularities cancel.

We have checked numerically that all spurious singularities do indeed 
cancel in the entire amplitude, for each of the four NMHV helicity
configurations.  This cancellation is a strong check on 
the amplitude, precisely because it involves the analytic behavior 
of the integral functions in a non-trivial way.


\section{An All-Multiplicity Coefficient}
\label{AllNSection}

\def\lsl{\not{\hbox{\kern-2.3pt $\ell$}}}

The computation of amplitudes beyond a fixed number of external
legs can provide additional information and clues to general structure.
In particular, as we shall discuss in \sect{TwistorSection}, a discussion
of the twistor-space structure of the seven-point amplitude requires
an accounting for the `holomorphic anomaly' pointed out by Cachazo,
Svr\v{c}ek, and Witten~\cite{CSWIII}.  Some aspects of the structure
of an all-$n$ amplitude also require such an accounting; but we will
be able to describe most aspects without reference to the `anomaly'.

We limit ourselves here to confirming the coefficient presented in
ref.~\cite{Cachazo}; presenting one new set of coefficients, of a
class of three-mass box functions; and discussing a number of
vanishing coefficients.

Consider the adjacent-minus amplitude $A_{n;1}(1^-,2^-,3^-,4^+,\ldots,n^+)$.
The simplest of the cuts is the cut in the $s_{123}$ channel, represented
by the product of tree amplitudes, 
\begin{equation}
 C_{123} \equiv  i
   \, A^\tree_5((-\ell_1)^+,1^-,2^-,3^-,\ell_4^+)
   \, A^{\tree}_5((-\ell_4)^-,4^+,5^+,\ldots n^+,\ell_1^-)\,,
\label{AllnCut1}
\end{equation}
integrated over phase space.   
This cut is particularly simple because only MHV (or 
$\overline{\hbox{MHV}})$ amplitudes 
participate, and only gluons can cross the cut (so the result
is the same in $\Neqfour$ or pure $\Neqone$ super-Yang-Mills theory,
or QCD).  Inserting~\eqn{PT} twice, the cut becomes
\begin{equation}
 C_{123} = 
{ i(s_{123})^3  \over \spb1.2\spb2.3\spa4.5\spa5.6 \cdots \spa{(n-1)}.{n} } \; 
   {1 \over \spb{\ell_1}.{1} \spb{3}.{\ell_4}
            \spa{\ell_4}.{4} \spa{n}.{\ell_1} } \,.
\label{AllnCut2}
\end{equation}
We denote the running loop momenta by $\ell_i$; that is, 
$\ell_n=\ell_1+k_n$, $\ell_1$, $\ell_2=\ell_1 - k_1$, 
$\ell_3=\ell_1 - k_1 - k_2$, $\ell_4=\ell_1 - k_1 - k_2 - k_3$. 
To obtain \eqn{AllnCut2} we have used 
$\spa{\ell_1}.{\ell_4} \spb{\ell_1}.{\ell_4} = (\ell_1-\ell_4)^2 = s_{123}$. 
We can also clear out the loop-momentum-dependent spinor-products
from the denominator, in favor of standard propagators,
\begin{equation}
 C_{123} = { i(s_{123})^3  \over \spb1.2\spb2.3\spa4.5\spa5.6 \cdots
       \spa{(n-1)}.{n} } \; 
 { \langle 3^- | \lsl_4 | 4^- \rangle \langle 1^- | \lsl_1 | n^- \rangle
    \over \ell_n^2 \ell_2^2 \ell_3^2 \ell_5^2  }  \,.
\label{AllnCut3}
\end{equation}
It turns out that the algebraic steps for integrating the
cut are essentially identical to those needed in the six-point
case~\cite{Neq1Oneloop}.  One can actually do a bit better,
and avoid any integration, by multiplying and dividing by the 
spinor strings,
\begin{equation}
\sand{n}.{(1 + 2)}.{3} \sand{4}.{(2 + 3)}.{1} \,,
\label{Closer}
\end{equation}
whose appearance is motivated by the form~(\ref{PlanarDefinition}) of the 
denominators encountered in reducing non-adjacent two-mass 
pentagon integrals.
One obtains,
\begin{eqnarray}
 C_{123} &=& { i(s_{123})^3  \over \spb1.2\spb2.3\spa4.5\spa5.6 \cdots
       \spa{(n-1)}.{n} \spba{1}.{(2 + 3)}.{4} \spba{3}.{(1 + 2)}.{n}}
\nonumber \\ &&\hskip0.5cm \times
 { {\rm tr}\Bigl[ \textstyle{{1\over2}} (1+\gamma_5) (1+2) 3 \lsl_4 4 
                         (2+3) 1 \lsl_1 n \Bigr]
    \over \ell_n^2 \ell_2^2 \ell_3^2 \ell_5^2  }  \,.
\label{AllnCut4}
\end{eqnarray}
The trace of Dirac $\gamma$ matrices, with a chiral projection
inserted, is easily evaluated. 
The $\gamma_5$ terms are proportional to the Levi-Civita tensor and 
integrate to zero. One can algebraically reduce the non-Levi-Civita 
terms to a combination of four cut box integrals,
\begin{eqnarray}
&& A^\oneloop(1^-,2^-,3^-, 4^+, \ldots, n^+) \Bigr|_{123\ \rm cut} \nonumber \\
&& \hskip 2 cm \null
 = i \cg \,
  { s_{123}^3 \over \spb1.2 \spb2.3 \spa4.5 \spa5.6 \cdots \spa{(n-1)}.{n} 
           \spba{1}.{(2 + 3)}.{4} \spba{3}.{(1 + 2)}.{n} } 
 \nonumber \\
& & \hskip 4 cm 
\times \Bigl( F^{\rm 1m}(s_{12}, s_{23}, s_{123})
       + F^{{\rm 2m}\,h}(s_{n1}, s_{123}, s_{23}, s_{4\ldots(n-1)}) \nonumber \\
&&\null \hskip 4.5 cm 
        + F^{{\rm 2m}\,h}(s_{34}, s_{123}, s_{5\ldots n}, s_{12}) \nonumber \\
&& \null \hskip 4.5 cm 
          + F^{{\rm 2m}\,e}(s_{n\ldots3}, s_{1\ldots4}, s_{123}, 
               s_{5\ldots(n-1)})  \Bigr) \,,
\label{C123Result}
\end{eqnarray}
where the integral functions $F$ are given in \app{IntegralsAppendix}.  
This result reduces to \eqns{mmmppppc567}{mmmppppc567s} for $n=7$.
It also confirms Cachazo's~\cite{Cachazo} 
very elegant evaluation of this coefficient, obtained by exploiting 
the holomorphic anomaly~\cite{CSWIII}.

\begin{figure}[t]
\centerline{\epsfxsize 2.5 truein \epsfbox{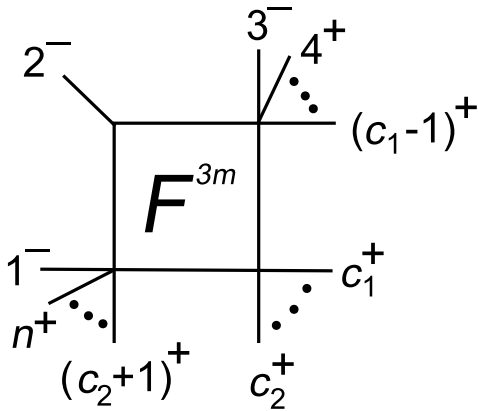}}
\caption[a]{\small The class of three-mass box functions whose
coefficient is given in \eqn{Alln3m}.}
\label{allnbox3mFigure}
\end{figure}

Now consider the coefficient of a class of three-mass box functions in the
same amplitude.  In this case NMHV tree amplitudes appear in the cuts.  
We used the simplified form of the NMHV amplitudes appropriate for use 
in the cuts, obtained in ref.~\cite{NMHVTree}.  
The cut starts out with the structure of a
tensor octagon integral (independent of $n$), and can be cleanly
reduced to a sum of scalar box integrals.  
Here we present the coefficient of the set of three-mass box integrals,
\begin{equation}
F^{3{\rm m}}(s_{2 \ldots (c_1-1)}, s_{(c_2+1)\ldots 2},
             s_{3\ldots(c_1-1)}, s_{c_1\ldots c_2}, s_{(c_2+1)\ldots 1})\,,
\label{Alln3mboxfunction}
\end{equation}
shown in \fig{allnbox3mFigure}.
This coefficient can be written as a single term,
\begin{eqnarray}
i \cg \, 
{ ( \spa{1}.{2} \spa{2}.{3} s_{c_1\ldots c_2} )^4
   \over\spa1.2\spa2.3\cdots\spa{n}.1 }
&\times& { \spa{(c_1-1)}.{c_1} \spa{c_2}.{(c_2+1)}
 \over 
    s_{c_1\ldots c_2} 
    \sandmp{(c_1-1)}.{\Ksl_{c_1\ldots c_2} \Ksl_{(c_2+1)\ldots 1}}.{2}
     \sandmp{c_1}.{\Ksl_{c_1\ldots c_2} \Ksl_{(c_2+1)\ldots 1}}.{2}} 
\nonumber \\
&\times& 
  {1 \over 
      \sandmp{c_2}.{\Ksl_{c_1\ldots c_2} \Ksl_{3\ldots (c_1-1)}}.{2}
      \sandmp{(c_2+1)}.{\Ksl_{c_1\ldots c_2} \Ksl_{3\ldots(c_1-1)}}.{2}}
     \,.
\nonumber \\
&& ~
\label{Alln3m}
\end{eqnarray}
We have checked that this set of coefficients correctly reproduces the
appropriate box coefficients for $n=6$ and 7.

We defer a more detailed discussion of this computation, as well as
results for other box coefficients, to a future publication.  While
the all-$n$ NMHV loop amplitude contains everything from one-mass to
three-mass boxes (four-mass boxes are absent), a generic term contains
either a three-mass box, or an easy two-mass box.  The latter coefficients
are more complicated in structure, while the former are simpler. 


Using the triple cut shown in \fig{triplecutFigure}, it is easy to
see that {\it the coefficient of any box integral which contains two
adjacent clusters of legs, each having only positive-helicity gluons,
must vanish.}   This statement is true for any amplitude in any
cut-constructible theory.   The reasoning is as follows:
Consider a triple cut where each of the two positive-helicity clusters
is identified with the set of external legs emerging from a blob
in \fig{triplecutFigure}.   Then the vanishing of the tree amplitudes
$A_m^{\tree}({\pm}{+}{+}\cdots{+})$ implies that the particle
crossing the cut line between these two blobs must have $(-)$ helicity
on both sides of the cut, which is not a valid helicity assignment.
A corollary is that {\it any box integral with one all-plus cluster
adjacent to two consecutive massless plus-helicity legs must
have a vanishing coefficient.}
A simple consequence of this result for the seven-point 
amplitude $({-}{-}{-}{+}{+}{+}{+})$ is that
$c_{235} = c_{237} = c_{257} = c_{357} = 0$, 
as given in \eqn{mmmppppczeroes}.
Clearly a large number of box coefficients for the all-$n$ amplitude
$({-}{-}{-}{+}{+}\cdots{+}{+})$ vanish by these rules.


\section{Twistor-Space Properties}
\label{TwistorSection}

\def\psl{\not{\hbox{\kern-2.3pt $p$}}}
\def\tlambda{\tilde\lambda}
\def\adot{{\dot a}}
\def\bdot{{\dot b}}
\def\tspa#1.#2{\left\langle#1,#2\right\rangle}
\def\tspb#1.#2{\left[#1,#2\right]}
The target space for Witten's candidate topological string theory is
$\CP^{3|4}$, otherwise called projective (super-)twistor space.  Points
in twistor space correspond to null momenta or equivalently to light cones
in space-time.  The correspondence is specified by a `half-Fourier'
transform.  More precisely, if we represent a null momentum by the
tensor product of a spinor $\lambda^a$ and a conjugate spinor 
$\tlambda^\adot$,
then twistor quantities are obtained by Fourier-transforming with respect
to all the $\tlambda^\adot$.

Amplitudes in twistor space, as it turns out, have rather simple
properties.  At tree level, they are non-vanishing only on certain curves.
This implies that they contain factors of delta functions (or derivatives
thereof) whose arguments are the characteristic equations for the curves.
The coefficients of the delta functions, however, have been quite
difficult to calculate directly.  

As Witten pointed out in his original
paper~\cite{WittenTopologicalString}, however, we do not need the
twistor-space amplitudes in order to establish the structure of the delta
functions they contain.  In momentum space, the Fourier transform turns
the polynomials into differential operators (polynomial in the
$\lambda_i$, and derivatives with respect to the $\tlambda_i$), which will
annihilate the amplitude.  One particularly useful building block for
these differential operators is the line annihilation operator, expressing
the condition that three points in twistor space lie on a common `line' or
$\CP^1$.  If the coordinates of the three points, labeled $i,j,k$, 
are $Z^I_i = (\lambda^a_i,\mu^{\adot}_i)$, {\it etc.}, then 
the appropriate condition is
\begin{equation}
\epsilon_{IJKL} Z^I_i Z^J_j Z^K_k = 0\,,
\end{equation}
for all choices of $L$.  Choosing $L=\adot$, and translating this
equation back to momentum space using the identification 
$\mu^\adot \leftrightarrow -i\partial/\partial\tlambda_\adot$, we 
obtain the operator,
\be
F_{ijk} = \spa{i}.{j} 
{\partial\over\partial\tlambda_k}
+\spa{j}.{k} 
{\partial\over\partial\tlambda_i}
+\spa{k}.{i} 
{\partial\over\partial\tlambda_j} \,.
\label{Fdef}
\ee

Two important sufficient conditions for $F_{ijk}$ to annihilate
an expression, {\it i.e.} for it to have support only when $i,j,k$ lie on 
a line in twistor space, are~\cite{WittenTopologicalString}
\begin{enumerate}
\item The expression is completely independent of $\tlambda_i$, $\tlambda_j$,
and $\tlambda_k$, or
\item $\tlambda_i$, $\tlambda_j$, $\tlambda_k$ appear only via a sum of 
momenta containing them, of the form 
\be
P^{a\adot} = (\cdots + k_i + k_j + k_k + \cdots)^{a\adot}
  = \cdots + \lambda_i^a\tlambda_i^\adot 
           + \lambda_j^a\tlambda_j^\adot
           + \lambda_k^a\tlambda_k^\adot
           + \cdots \,.
\label{ijkMomSum}
\ee
\end{enumerate}
The first condition is obvious from the definition~(\ref{Fdef}); 
the second holds because of the Schouten
identity,
\be
  \spa{i}.{j} \lambda_k 
+ \spa{j}.{k} \lambda_i
+ \spa{k}.{i} \lambda_j = 0\,.
\label{Schouten}
\ee

The tree-level MHV amplitude, for example, is annihilated by $F_{ijk}$,
because it is independent of the $\tlambda_i$.  Any possible delta
functions vanish for generic momenta, because they take the form
$\delta(\spa{i}.{j})$.  At one loop, 
Cachazo, Svr\v{c}ek, and Witten~\cite{CSWIII} pointed out that
such delta
functions, arising from the spinor analog of the fact that 
$\partial_{\overline z}\, (1/z)\neq 0$, do arise.  They must be taken
into account for a proper analysis of the twistor-space structure
of amplitudes.

We will not compute the relevant `holomorphic anomaly' terms for the amplitudes
in this paper, and so we will not be able to fully exhibit their
twistor-space structure.  While the `anomaly' terms enter into the
action of the differential operators on the box integrals, their
action on the coefficients is unaffected by it.  The properties
of the coefficient are also important, and we discuss them.

In addition to the line operator $F_{ijk}$, we will employ the 
planar operator~\cite{WittenTopologicalString},
\be
K_{ijkl} \equiv \epsilon_{IJKL} Z^I_i Z^J_j Z^K_k Z^L_l
 = \spa{i}.{j} \e^{\adot\bdot}
   {\partial\over\partial\tlambda_k^\adot}
   {\partial\over\partial\tlambda_l^\bdot}
  \pm \hbox{[5 permutations]} \,,
\label{Kdef}
\ee
whose vanishing implies that four points lie in a plane in twistor space.

In the case of the seven-point amplitude, all line operators $F$
and planar operators $K$ are affected by the `anomaly'.  For higher-point
amplitudes, however, there are classes of unaffected operators.  We
will study their action on the term in the amplitude
discussed in \sect{AllNSection}.  We shall show it is consistent
with the simple twistor-space structure expected from a generalization
of the BST calculation.


\subsection{Properties of seven-point box coefficients $c_{ijk}$}
\label{SevenptCSubsection}

Before describing the twistor-space structure of the coefficients
$c_{ijk}$ from section~\ref{ResultsSection}, we list some
of their other (related) properties here.
\begin{itemize}
\item Each box coefficient can be written as the sum of a small number
of `terms', or `one-term coefficients':
namely, the auxiliary quantities $c_A,c_B,c_C,c_D,c_E$ (as needed), 
plus the $c_{ijk}$ that are given directly in
section~\ref{ResultsSection} in terms of spinor strings.
\item The sum contains at most four terms.  In the case of 
$A_{7;1}^{\Neqfour}(1^-,2^-,3^-,4^+,5^+,6^+,7^+)$, it has at most three terms.   
\item Only the coefficients of the one-mass box integrals, $c_{i,i+1,i+2}$
(with indices mod 7), require the use of three or four terms.  
\item The coefficients of the hard two-mass box integrals, $c_{i,i+1,i+3}$
or $c_{i-2,i,i+1}$, require at most two terms.
\item The coefficients of the easy two-mass box integrals, $c_{i,i+1,i+4}$,
and of the three-mass box integrals, $c_{i,i+2,i+4}$, are all one-term
expressions.
\item Each term resembles a product of off-shell MHV vertices~\cite{CSW},
in terms of the types of denominators that occur: 
multi-particle invariants $s_{i,i+1,i+2}$, but not full two-particle poles
$s_{i,i+1}$; spinor products continued `off-shell', like 
$\spa{a}.{B^*} = \spab{a}.{(b+c)}.{d}$, where $k_B=k_b+k_c$, and $d$
temporarily plays the role of the arbitrary reference spinor in the CSW
formalism (see also ref.~\cite{NMHVTree}).  
The choice of $d$ varies from term to term, however.  
The longer strings of the form $\spaa{a}.{(b+c)}.{(d+e)}.{f}$
could correspond to continuing both spinors off-shell.
\item The resemblance extends to the numerator factors, which
always appear raised to the third or fourth power, like the factor
of $\spa{i}.{j}^4$ in the numerator of the Parke-Taylor
amplitudes~\cite{ParkeTaylor}.  Probably they 
should always be considered raised to the fourth power, 
because whenever such a factor appears raised to the third power, 
it always seems to be of the type that `could have appeared' 
in the denominator, such as
$\spa{i,}.{i+1}$, $\spb{i,}.{i+1}$, $s_{i,i+1,i+2}$, or even the 
more complicated spinor strings which are `square roots' of the integral
reduction factors discussed in section~\ref{CalculationSection}.
Other numerator factors which cannot occur in denominators, such as
$\spa{i,}.{i+2}$, or $\spab{a}.{(b+c)}.{d}$ where $a,b,c,d$ are {\it not}
cyclicly consecutive, always appear raised to the fourth power,
if they appear in the numerator at all.
\item Perhaps the most interesting of the one-term coefficients are those
exemplified by $c_{135}$ in $A_{7;1}^{\Neqfour}(1^-,2^-,3^+,4^-,5^+,6^+,7^+)$. 
Here a new structure appears in the numerator, unlike any of
the denominator factors.  The unexpected form made the simplified form of
this coefficient 
more difficult to guess.  In the end, we deduced it from its 
collinear behavior in particular channels.
Using the Schouten identity and momentum conservation,
one can show that it has the following collinear limits,
\bea
\hskip-0.5cm
\spba3.{(6+7)}.1 \spa4.6 + \spba3.{5}.4 \spa1.6 
&\to& + \spba3.{(6+7)}.1 \spa4.6 \hbox{   for $4 || 5$,} 
\label{newstrcoll45} \\
&\to& - \spba3.{(2+4)}.1 \spa4.6 \hbox{   for $5 || 6$,} 
\label{newstrcoll56} \\
&\to& - \spba3.{(1+2)}.4 \spa1.6 \hbox{   for $6 || 7$,} 
\label{newstrcoll67} \\
&\to& + \spba3.{(5+6)}.4 \spa1.6 \hbox{   for $7 || 1$.} 
\hskip.8cm
\label{newstrcoll71}
\eea
\end{itemize}

The most general twistor-space property of the box coefficients
is that {\it all points lie in a plane}.  That is, $K_{mnpq}$ for
every choice of $m,n,p,q$ annihilates every one-term coefficient,
and hence, by linearity, it annihilates every box coefficient $c_{ijk}$.
This property is not so easy to see analytically, but it is
straightforward to verify numerically by evaluating the expressions
$K_{mnpq}\ c_{ijk}$ at random points in 7-particle phase space.

\begin{figure}[t]
\centerline{\epsfxsize 6.0 truein \epsfbox{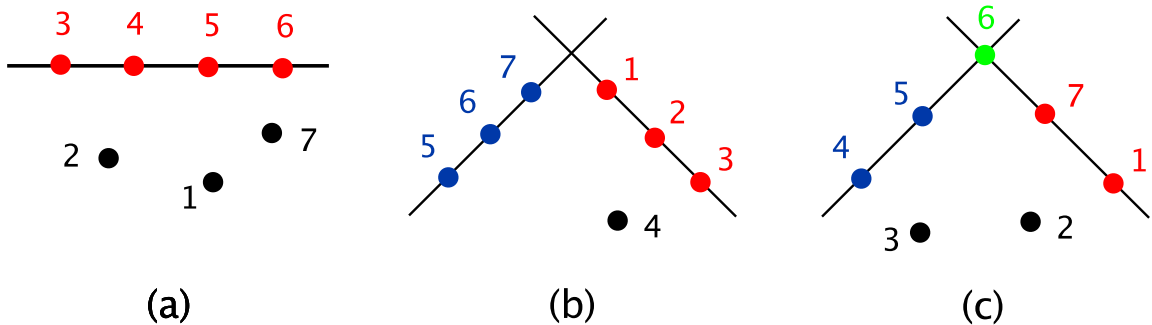}}
\caption[a]{\small Examples of twistor-space configurations for
single-term box coefficients in the helicity amplitude
$A_{7;1}^{\Neqfour}(1^-,2^-,3^+,4^-,5^+,6^+,7^+)$.
In every case, all the points lie in a plane.
(a) the easy two-mass box coefficient $c_{125}$, 
(b) the hard two-mass box coefficient $c_{237}$,
(c) the three-mass box coefficient $c_{135}$.}
\label{SampleConfigsFigure}
\end{figure}

On the other hand, the way the 7 coplanar points get distributed 
into lines varies from term to term, yet it does so in a very
systematic way.  Also, the correct behavior can be determined by inspection.
From the amplitude $A_{7;1}^{\Neqfour}(1^-,2^-,3^+,4^-,5^+,6^+,7^+)$,
consider, for example:
\begin{itemize}
\item the easy two-mass box coefficient $c_{125}$ given in
\eqn{mmpmpppc125}.  Here legs 3,4,5,6 only appear as $\lambda_i$
spinors, not $\tlambda_i$ spinors.  So legs 3,4,5,6 must lie on a line
for $c_{125}$.  On the other hand, legs 7,1,2
appear sometimes as $\tlambda_i$ spinors, as in $\spb1.2$ and $\spb7.1$,
so they should be placed at generic points in the plane, as shown
in \fig{SampleConfigsFigure}a.
\item the hard two-mass box coefficient $c_{237}$ given in
\eqn{mmpmpppc237}.  Here legs 1,2,3 either appear as $\lambda_i$
spinors, or through the momentum sum $k_1+k_2+k_3$.
The latter appearance is easy to see in $s_{123} = (k_1+k_2+k_3)^2$, 
but it is also true for 
$\spba4.{(2+3)}.1 = \spba4.{\gamma^\mu}.1 (k_1+k_2+k_3)^\mu$,
since $\psl_1 | 1^+ \rangle = 0$; and
similarly for $\spaa3.{(1+2)}.{(6+7)}.5$.
Thus legs 1,2,3 lie on a line.  Likewise, legs 5,6,7 lie on a different
line.  Leg 4 appears as $\tlambda_4$, so it sits in the plane, away
from the 2 lines, as shown in \fig{SampleConfigsFigure}b.
\item the three-mass box coefficient $c_{135}$ given in
\eqn{mmpmpppc135}.  From the denominator factors alone, one would 
conclude, as in the previous case that legs 4,5,6 lie in a line,
and legs 6,7,1 lie in a line.  Thus the two lines intersect at a common
point, leg 6.  The form of the numerator in \eqn{mmpmpppc135}
makes manifest the proper 6,7,1 behavior.  However, it is not apparent
that it allows 4,5,6 to lie in a line until one rewrites
\be
  \spba3.{(6+7)}.1 \spa4.6 + \spba3.{5}.4 \spa1.6
= \spba3.{(5+6)}.4 \spa1.6 + \spba3.{7}.1 \spa4.6 \,.
\label{c135numerid}
\ee
\end{itemize}
Numerical investigation reveals that the pictures shown 
in~\fig{SampleConfigsFigure} are accurate, in that there are no
`hidden' twistor-space co-linearities.
Note also that each configuration in~\fig{SampleConfigsFigure}
can be described as follows:  {\it all seven points lie in a plane, 
distributed among three lines, and two of the lines can be chosen 
to intersect at one of the seven points.}

In fact, the configurations shown in~\fig{SampleConfigsFigure},
after permuting the external leg labels around,
are the complete set of twistor-space configurations
for {\it all} the one-term coefficients for all four helicity
amplitudes.  Furthermore,
\begin{itemize}
\item The three-mass box coefficients (always one term) always 
have the configuration of~\fig{SampleConfigsFigure}c.
The massless leg in the corresponding integral always lies at the 
intersection of the two lines.
\item The easy two-mass box coefficients (always one term) always 
have the configuration of~\fig{SampleConfigsFigure}a.
The three points lying off the line in \fig{SampleConfigsFigure}a
are those belonging to the three-leg cluster forming one of the external
masses in the integral.
\item The hard two-mass box coefficients 
are the sums of (at most two) terms with the configurations 
of~\fig{SampleConfigsFigure}a and~\fig{SampleConfigsFigure}b.
In the \fig{SampleConfigsFigure}a term, the three points lying off 
the line are those belonging to the two-leg cluster and the 
massless leg adjacent to it.  In the \fig{SampleConfigsFigure}b term,
the point lying off of both lines is the massless leg adjacent to
the three-leg cluster.
\item The one-mass box coefficients generally involve all three
types of configurations in \fig{SampleConfigsFigure}.
However, for all helicity amplitudes the full coefficients do obey
$F_{712} F_{456} c_{123} = 0$ and $F_{123} F_{456} c_{123} = 0$,
plus all cyclic permutations thereof.
\end{itemize}

The coefficients of the box integral functions in the six-point NMHV
amplitude~\cite{Neq1Oneloop} have an analogous structure,
only simpler because there are fewer points:
All six points lie in a plane in twistor space, and for each single 
term in a box coefficient, three of the points always lie on a line.


\subsection{Properties of an all-$n$ box coefficient}
\label{AllnCSubsection}

\begin{figure}[t]
\centerline{\epsfxsize 3.5 truein \epsfbox{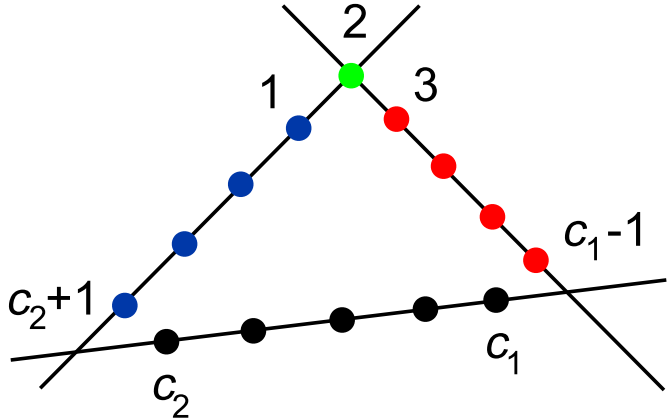}}
\caption[a]{\small The twistor-space configuration for
the class of three-mass-box coefficients in the helicity amplitude
$({-}{-}{-}{+}{+}\cdots{+}{+})$ discussed in the text.  All points
lie in a plane.}
\label{allntwistFigure}
\end{figure}

The result for the class of three-mass-box coefficients for the 
all-$n$ NMHV amplitude $({-}{-}{-}{+}{+}\cdots{+}{+})$ 
is given in \eqn{Alln3m}.  We now analyze its twistor-space structure.
First observe that all points in the set $\{ c_1,c_1+1,\ldots,c_2 \}$ 
(where the indices are understood cyclicly $\mod n$) lie on a line,
because they only appear through the momentum sum $K_{c_1\ldots c_2}$
or through $\lambda_{c_1}$ or $\lambda_{c_2}$.
The points $\{ c_2+1,c_2+2,\ldots,1,2 \}$ also lie on a line,
for a similar reason.  (Note that 
$ \Ksl_{(c_2+1) \ldots 1} | 2^+\rangle 
= \Ksl_{(c_2+1) \ldots 2} | 2^+ \rangle$,
and that leg 2 otherwise only appears as $\lambda_2$.)
Finally, the points $\{ 2,3,\ldots,c_1-1 \}$ lie on a line.
Numerical investigation for a number of randomly-chosen
box coefficients for $n=8,9,10$ confirms that all $n$ points lie in a plane.
This twistor-space structure is summarized in \fig{allntwistFigure}.

This structure generalizes the behavior of the seven-point `one-term'
coefficients, which was illustrated in \fig{SampleConfigsFigure}.
Again a `one-term' expression can be described in twistor space as
having all points lying on three lines within a single plane; and  
one of the intersections of the lines always contains
one of the $n$ points.

In the full amplitude, the coefficient~(\ref{Alln3m}) appears
multiplied by the three-mass-box integral function $F^{3{\rm m}}(s_{2
\ldots (c_1-1)}, s_{(c_2+1)\ldots 2}, s_{3\ldots(c_1-1)}, s_{c_1\ldots
c_2}, s_{c_2+1\ldots 1})$ in~\eqn{Alln3mboxfunction}.  The action of
the $F$ and $K$ twistor-space operators on $F^{3{\rm m}}$ is affected
by the holomorphic anomaly.  The behavior of leg 2 in particular is
affected, because of singularities in the integral when the loop
momentum becomes collinear with $k_2$.  At the level of the $F$
operator, none of the other legs are affected by this issue, since
they are safely buried into massive clusters.  So, the {\it lines} in
\fig{allntwistFigure}, excluding the point 2 and ignoring their
intersections, also describe the twistor-space collinear behavior of
the box function multiplying the coefficient.  (This property is also
manifest from the form of the arguments of $F^{3{\rm m}}$.)  On the
other hand, the co-planarity of (the finite parts of) this integral is
not apparent, and perhaps doubtful, even taking into account the
holomorphic anomaly.


\section{Conclusions}
\label{ConclusionsSection}

In this paper we computed all the next-to-MHV one-loop seven-gluon
amplitudes in $\NeqFour$ super-Yang-Mills theory, as well as a
class of contributions to an $n$-gluon next-to-MHV amplitude.
These results provide a useful guide to the spinor-product structure 
and (related) twistor-space properties of higher-point $\NeqFour$ amplitudes.
The coefficients of the box integral functions appearing in the amplitudes
can be written as simple terms built out of spinor strings, or sums
of just a few such terms.  Each term exhibits a simple twistor-space 
structure:  all points lie on three lines confined to a single plane, 
and two of the lines always intersect at one of the $n$ points.

The co-planarity of the NMHV box coefficients is a very intriguing
result.  This complete co-planarity may be demonstrated for general
one-loop NMHV amplitudes along the same lines used by Cachazo to
demonstrate a certain degree of co-linearity~\cite{Cachazo}.  Using
the co-linearity of MHV tree amplitudes, Cachazo showed that $F_{ijk}$
annihilates a box coefficient if the points $i,j,k$ belong to the MHV
side of a `nontrivial' cut, one for which the corresponding box
function has an imaginary part. If the coefficient were not
annihilated, logarithms would remain in the cut after applying
$F_{ijk}$, in contradiction to the result obtain by integrating the
holomorphic anomaly in the cut~\cite{BBKRTwistor,Cachazo}.  (Because
the anomaly delta function freezes the phase-space integration, the
action of $F_{ijk}$ can give only a rational function or zero.)  The
same logic can be used to determine all $K_{ijkl}$ and $F_{ijk}$ that
annihilate a given box coefficient.  On one side of a cut the tree
amplitude is MHV and therefore supported on a line.  On the other side
of the cut the tree amplitude is NMHV so it is supported on pairs of
intersecting lines.  From the lack of logarithms in the integrated
anomaly, this implies that the coefficient of any integral with a cut
in a given channel satisfies the same coplanarity and collinearity 
requirements satisfied by the tree amplitudes on either side of the cut.
The complete co-planarity of a box coefficient can be demonstrated by
combining the information from overlapping cuts of the associated
integral.  This simple argument works for all coefficients, except for
those for one-mass boxes, or for easy two-mass boxes when one of the
masses is composed of only two external legs.  In these cases,
however, one can appeal to the linear equations obtained from the
infrared singularities, \eqn{OneLoopIRPoles}, for two- and
three-particle invariants, in order to argue (for $n>6$) that these
coefficients must also be co-planar.

This twistor-space co-planarity is stronger than would seem to be
implied by generic MHV loop diagrams.  For NMHV one-loop amplitudes
these diagrams have three vertices.  The one-particle-irreducible
MHV-diagrams (triangle diagrams) heuristically appear to be planar.
(An MHV vertex is dual to a line in twistor space; a propagator in the
loop is dual to an intersection between two
lines~\cite{WittenTopologicalString,CSW,CSWII}.)  On the other hand,
the one-particle-reducible MHV-diagrams (bubbles with an additional
MHV vertex attached to an external leg) do not appear to have to be
planar.  Apparently the sum is better behaved than individual MHV
diagrams.

The distribution of the seven points into lines in the plane, for the
one-term coefficients in $A_{7;1}^{\NeqFour}$, as depicted in
\fig{SampleConfigsFigure}, is also intriguing.  This division is
independent of the particular NMHV helicity configuration, and only
depends on the type of integral being considered.  This
helicity-independence seems obscure from the point of view of MHV
diagrams.

The simplicity of the structure we have uncovered suggests that the 
computation of further multi-leg one-loop $\NeqFour$ amplitudes will 
be quite tractable, and provides strong motivation for additional 
investigations in this direction.

The seven-point results presented here may also have some practical
relevance.  They can be used to benchmark purely numerical approaches to
complicated multi-parton loop amplitudes in QCD, which are likely to
become available in the near future.  Such amplitudes will contribute to a
better understanding of cross sections for multi-jet production at hadron
colliders.  Multi-jet rates are important to understand for their own
sake, but also because such processes form both irreducible and reducible
backgrounds (the latter when jets fake leptons or photons, for example) to
processes involving the electroweak interactions.  
We are optimistic that combining insights from unitarity and twistor space
will provide new practical tools for computing experimentally-relevant
loop amplitudes in massless gauge theories.


\section*{Acknowledgments}

We thank the KITP at Santa Barbara, where part of this work was
carried out, for its hospitality.  V.~D.~D., L.~J.~D. and D.~A.~K. are
also grateful to the IPPP at Durham for its hospitality during part of
the research.  We are grateful to Edward Witten for
motivating this project, and to Iosif Bena, Niels Emil Bjerrum-Bohr,
David Dunbar, Radu Roiban, Marcus Spradlin and Anastasia Volovich for
helpful discussions.  We also thank Academic Technology Services at
UCLA for computer support. Some of the diagrams in this paper were
constructed with {\tt JaxoDraw}~\cite{Jaxodraw}.


\appendix

\section{Box Integrals}
\label{IntegralsAppendix}

In this appendix we collect the dimensionally-regulated integral
functions appearing in the $\NeqFour$ amplitudes; these integral
functions were obtained from refs.~\cite{FourMassBox,Integrals5}.
The reader is referred to these papers for further details of their
derivation.  Through $\Ord(\eps^0)$, we have 
\def\hs{\hskip 2.5 cm \null}
\begin{eqnarray}
&& F^{4{\rm m}}  (s, t, K_1^2, K_2^2, K_3^2, K_4^2)
= {1\over 2}
\Biggl\{ - \Li_2\left(\hf(1-\lambda_1+\lambda_2+\rho)\right)
   + \Li_2\left(\hf(1-\lambda_1+\lambda_2-\rho)\right) \nonumber \\
 && \hs
  - \Li_2\left(
   \textstyle-{1\over2\lambda_1}(1-\lambda_1-\lambda_2-\rho)\right)
  + \Li_2\left(\textstyle-{1\over2\lambda_1}(1-\lambda_1-
    \lambda_2+\rho)\right) \nonumber \\
 && \hs - {1\over2}\ln\left({\lambda_1\over\lambda_2^2}\right)
   \ln\left({ 1+\lambda_1-\lambda_2+\rho \over 1+\lambda_1
        -\lambda_2-\rho }\right) \Biggr\} \,, 
  \label{Fboxes4m} \\
&&  F^{3{\rm m}}(s,t, K_2^2, K_3^2, K_4^2)
=  -{1\over2\e^2} \Bigl[ (-s)^{-\e} + (-t)^{-\e}
     - (-K_2^2)^{-\e} - (-K_4^2)^{-\e} \Bigr]\nonumber \\
&& \hs    - {1\over2} \ln\left({-K_2^2\over -t}\right)
                \ln\left({-K_3^2\over -t}\right)
          - {1\over2} \ln\left({-K_3^2\over -s}\right)
                \ln\left({-K_4^2\over -s}\right)  \nonumber\\
 &&\hs + \Li_2\left(1-{K_2^2\over s}\right)
   + \Li_2\left(1-{K_4^2\over t}\right) 
   -  \Li_2\left(1-{K_2^2K_4^2\over st}\right) \nonumber\\
 &&\hs + {1\over 2} \ln^2\left({-s\over -t}\right) , 
  \label{Fboxes3m} \\
&&  F^{2{\rm m} \, h} (s, t, K_3^2, K_4^2) 
=  -{1\over 2\e^2} \Bigl[ (-s)^{-\e} + 2 (-t)^{-\e}
              - (-K_3^2)^{-\e} - (-K_4^2)^{-\e} \Bigr]
\nonumber\\
&&\hs - {1\over2} \ln\left({-K_3^2\over -s}\right)
                \ln\left({-K_4^2\over -s}\right)
   + \Li_2\left(1-{K_3^2\over t}\right)
   + \Li_2\left(1-{K_4^2\over t}\right)  \nonumber\\
 &&\hs + {1\over 2} \ln^2\left({-s\over -t}\right) ,
 \label{Fboxes2mh} \\
&&  F^{2{\rm m}\, e}(s, t, K_2^2, K_4^2) 
=   -{1\over\e^2} \Bigl[ (-s)^{-\e} + (-t)^{-\e}
              - (-K_2^2)^{-\e} - (-K_4^2)^{-\e} \Bigr] \nonumber\\
 &&\hs  + \Li_2\left(1-{K_2^2\over s}\right)
    + \Li_2\left(1-{K_2^2\over t}\right)
    + \Li_2\left(1-{K_4^2\over s}\right)
    + \Li_2\left(1-{K_4^2\over t}\right) \nonumber\\
 &&\hs  - \Li_2\left(1-{K_2^2K_4^2\over st}\right)
    + {1\over2} \ln^2\left({-s\over -t}\right) ,
   \label{Fboxes2me} \\
&&  F^{1{\rm m}} (s,t, K_4^2) 
=  -{1\over\e^2} \Bigl[ (-s)^{-\e} + (-t)^{-\e} - (-K_4^2)^{-\e} \Bigr]
            \nonumber\\
&& \hs  + \Li_2\left(1-{K_4^2\over s}\right)
   + \Li_2\left(1-{K_4^2\over t}\right)
   + {1\over 2} \ln^2\left({-s\over -t}\right) + {\pi^2\over6} \, , 
  \label{Fboxes1m} \\
&&  F^{0{\rm m}}(s,t)  = 
 - {1\over\e^2} \Bigl[ (-s)^{-\e} + (-t)^{-\e} \Bigr]
  + {1\over 2} \ln^2\left({-s\over -t}\right) + {\pi^2\over 2} \,,
  \label{Fboxes0m} 
\end{eqnarray}
where the $k_i$ denote massless  momenta and the $K_i$ massive momenta.
The external momentum arguments $K_1,\ldots,K_4$ are sums
of external momenta $k_i$ from the $n$-point amplitude.  
The kinematic variables appearing in the integrals are 
\begin{equation}
s = (k_1 + k_2)^2 \, , \hs t = (k_2 + k_3)^2\,,
\end{equation}
or with $k$ relabeled as $K$ for off-shell (massive) legs.
The functions appearing in $F_4^{4 \rm m}$ are
\begin{equation}
 \rho\ \equiv\ \sqrt{1 - 2\lambda_1 - 2\lambda_2
+ \lambda_1^2 - 2\lambda_1\lambda_2 + \lambda_2^2}\ ,
\label{rdefinition}
\end{equation}
and
\begin{equation}
\lambda_1 = {K_1^2 \, K_3^2 
\over (K_1 + K_2)^2 \, (K_2 + K_3)^2 } \; , \hskip 1.5 cm
\lambda_2 = {K_2^2  \, K_4^2 \over
 (K_1 + K_2)^2 \, (K_2 + K_3)^2  } \; .
\end{equation}
We have rearranged the expressions for $F^{3{\rm m}}$
and $F^{2{\rm m} \, h}$ to make the poles in $\e$ 
more transparent.  We have also corrected some signs
in $F^{4{\rm m}}$ in ref.~\cite{Neq4Oneloop} and in the
published version of ref.~\cite{Integrals5}.


\section{ Multi-Particle Factorization Example}
\label{FactorizationAppendix}

In this appendix we explicitly evaluate the multi-particle
factorization of an NMHV seven-point amplitude in $\NeqFour$
super-Yang-Mills theory.  As discussed in the text,
one-loop multi-particle factorization is described by a universal
formula (\ref{LoopFact}), even though naive factorization does not
hold for infrared-divergent gauge theories.

As an example, consider the amplitude 
$A_{7;1}^{\NeqFour}(1^-, 2^-, 3^-, 4^+, 5^+, 6^+, 7^+)$.  
According to the general factorization formula, in the limit 
$K^2 = s_{234} \rightarrow 0$, where $K = k_2 + k_3 + k_4$, we have
\begin{eqnarray}
A_{7;1}^{\NeqFour}\
&& \hskip -.42 cm 
 \mathop{\longrightarrow}^{s_{234} \rightarrow 0} 
\hskip .15 cm 
      A_{4;1}^{\NeqFour}(2^-,3^-,4^+, (-K)^+) \, {i \over s_{234}} \, 
      A_{5}^{\tree}(K^-, 5^+, 6^+, 7^+, 1^-)
\nonumber \\
&&  \null \hskip .8 cm 
   + A_{4}^{\tree}(2^-,3^-,4^+, (-K)^+) \, {i \over s_{234}} \, 
      A_{5;1}^{\NeqFour}(K^-, 5^+, 6^+, 7^+, 1^-)
\label{LoopFact234} \\
&&\null \hskip .8 cm 
   +  A_{4}^{\tree}(2^-,3^-,4^+, (-K)^+) \, {i \over s_{234}} \, 
      A_{5}^{\tree}(K^-, 5^+, 6^+, 7^+, 1^-) \nonumber \\
&& \hskip 4 cm \null \times 
   \cg \, \Fact_7^{\NeqFour}(s_{234};k_1, \ldots, k_7) \,.
\nonumber
\end{eqnarray}
The subtlety in loop-level multi-particle factorization can 
be exposed by inspecting the infrared divergences 
on both sides of \eqn{LoopFact234}.
On the left-hand side, the divergences are~\cite{UniversalIR},
\begin{equation}
A_{7;1}^{\NeqFour}(1,2,3,4,5,6,7)\Bigr|_{\e\ {\rm pole}}\ =\
 -{\cg\over\e^2} \sum_{j=1}^7
         \left( { \mu^2 \over -s_{j,j+1} } \right)^\e 
 A_{7}^{\tree}(1,2,3,4,5,6,7) \ . 
\label{SevenPtSingular}
\end{equation}
On the other hand, in the products of amplitudes appearing in the
first two terms on the right-hand side of \eqn{LoopFact234}, we have
the singular terms
\begin{eqnarray}
&&  -{\cg\over\e^2} A_{7}^{\tree}(1,2,3,4,5,6,7) \Biggl[  
        \biggl( { \mu^2 \over s_{K2}} \biggr)^\e
      + \biggl( { \mu^2 \over -s_{23}} \biggr)^\e
      + \biggl( { \mu^2 \over -s_{34}} \biggr)^\e
      + \biggl( { \mu^2 \over s_{4K}} \biggr)^\e \nonumber \\
&& \hskip 2.0 cm
      + \biggl( { \mu^2 \over -s_{K5}} \biggr)^\e 
      + \biggl( { \mu^2 \over -s_{56}} \biggr)^\e
      + \biggl( { \mu^2 \over -s_{67}} \biggr)^\e
      + \biggl( { \mu^2 \over -s_{71}} \biggr)^\e
      + \biggl( { \mu^2 \over -s_{1K}} \biggr)^\e
\Biggr]
  \Biggr|_{s_{234} \rightarrow 0}. \hskip 1.7 cm 
\label{FactSevenPtSingular}
\end{eqnarray}
The mismatch between \eqn{SevenPtSingular} and
\eqn{FactSevenPtSingular} must be absorbed into the 
factorization function $\Fact_7$ appearing in the third term in
\eqn{LoopFact234}.  This mismatch implies, after expanding in $\e$,
that the factorization function must contain the logarithmic terms
\begin{equation}
 {\ln(-s_{12}) \over \e} + {\ln(-s_{45}) \over \e}\,.
\label{IRDiscontinuity}
\end{equation}
Neither of these terms allows for a separation of momenta 
$\{k_2, k_3, k_4\}$ from $\{ k_5, k_6, k_7, k_1 \}$, which one might 
have expected from a naive interpretation of \eqn{LoopFact234}, 
or from \fig{MultiFactFigure} with $i=2$ and $r=3$.

The mismatched infrared singularities in \eqn{IRDiscontinuity} encode
the factorization functions.  From \eqn{GeneralFactorizationFunction},
the infrared singularities~(\ref{IRDiscontinuity}) imply that the
factorization function in this channel is (recall 
$\Fact_n^{\NeqFour} = \Fact_n^{\rm non\hbox{-}fact}$),
\begin{eqnarray}
&& \Fact_7^{\rm non\hbox{-}fact}(s_{234}; k_1, \ldots, k_7) \nonumber\\
&& \null \hskip 1.5 cm  
 = 2 \, (\mu^2)^\eps \biggl(
     F^{{\rm 2m}\,h}(s_{12}, s_{234}, s_{34}, s_{567})
   + F^{{\rm 2m}\,h}(s_{45}, s_{234}, s_{671}, s_{23})
   + {2 \over \eps^2} (-s_{234})^{-\eps} \biggr) \nonumber \\
&& \null \hskip 1.5 cm  
= 
2 (\mu^2)^\eps \Biggl\{
 -{1\over\e^2} \biggl[ (- s_{12})^{-\e}
              - (-s_{567})^{-\e} - (-s_{34})^{-\e} \biggr] \nonumber \\
& & \hskip 2.8 cm \null 
 - {1\over2\e^2}
    {(-s_{567})^{-\e}(-s_{34})^{-\e} \over (- s_{12})^{-\e} }
   + {1\over 2} \ln^2\left({-s_{12}\over -s_{234} }\right) \nonumber\\ 
& & \hskip 2.8 cm \null 
   - {1\over 2} \ln^2 \left( {-s_{34} \over -s_{234}} \right)
   - {1\over 2} \ln^2 \left( {-s_{567} \over -s_{234}} \right) 
   - {\pi^2 \over 3} \Biggr\} \nonumber \\
& & \hskip 2.8 cm \null 
+ \{k_1 \leftrightarrow k_5, k_2 \leftrightarrow k_4, 
          k_6 \leftrightarrow k_7 \} \, .\hskip 2.5 cm \null
\label{FactFunc7}
\end{eqnarray}

Now compare this prediction to the results for
$A_{7;1}^{\NeqFour}(1^-,2^-,3^-,4^+,5^+,6^+,7^+)$ 
given in \sect{ResultsSection}.
From that section, eight of the box functions have an $s_{234}$ kinematic
pole in their coefficients. Thus,
\begin{eqnarray}
&& A_{7;1}^{\NeqFour}
    (1^-,2^-,3^-,4^+,5^+,6^+,7^+)\Bigr|_{s_{234} \rightarrow 0}
  \nonumber\\
&& \hskip 2.5 cm 
\longrightarrow\ c_{134} \, B(1,3,4) + c_{137} \, B(1,3,7)
               + c_{167} \, B(1,6,7) + c_{234} \, B(2,3,4)
  \hskip 1 cm \nonumber \\
&& \hskip 3. cm \null
               + c_{345} \, B(3,4,5) + c_{346} \, B(3,4,6) 
               + c_{347} \, B(3,4,7) + c_{467} \, B(4,6,7) 
 \,. \hskip 1 cm 
\label{T234PoleInts}
\end{eqnarray}
In this limit the coefficients appearing with an $s_{234}$ pole 
behave simply.  Five of the coefficients give,
\begin{equation}
\{ c_{134}, c_{234}, c_{345}, c_{346}, c_{347} \} \rightarrow
   A_4^\tree(2^-, 3^-, 4^+, (-K)^+) {1 \over s_{234}}
   A_5^\tree(K^-, 5^+, 6^+, 7^+, 1^-)  \,,
\label{FirstFiveMulti}
\end{equation}
while the remaining three give,
\begin{equation}
\{ c_{137}, c_{167}, c_{467} \} \rightarrow
 2 A_4^\tree(2^-, 3^-, 4^+, (-K)^+)\, {1 \over s_{234}}\,
    A_4^\tree(K^-, 5^+, 6^+, 7^+, 1^-) \,.
\label{RemainingThreeMulti}
\end{equation}

In the $s_{234} \rightarrow 0$ factorization limit, an inspection 
of the explicit form of the integrals in \app{IntegralsAppendix}
reveals, after expanding in $\e$, that the ones with an $s_{234}$ pole 
in their coefficients behave as 
(see also table~5 of ref.~\cite{BernChalmers}),
\begin{eqnarray}
&& B(2,3,4) = F^{{\rm 1m}} (s_{56}, s_{67}, s_{567})
    \longrightarrow 
  F^{{\rm 1m}} (s_{56}, s_{67}, s_{567}) \,,
\\
&& B(3,4,5) = F^{{\rm 1m}} (s_{67}, s_{71}, s_{671}) 
    \longrightarrow 
  F^{{\rm 1m}} (s_{67}, s_{71}, s_{671}) \,,
\\
&& B(1,3,4) = F^{{\rm 2m}\,h} (s_{56}, s_{671}, s_{71}, s_{234}) 
    \longrightarrow 
  F^{{\rm 1m}} (s_{671},s_{56}, s_{71})
   + {1\over \e^2} (-s_{234})^{-\e} \nonumber 
\\
   && \hskip 6.5 cm \null
   - {1 \over 2 \e^2} {(-s_{234})^{-\e} (-s_{71})^{-\e} \over (-s_{56})^{-\e}}
   - \Li_2\biggl(1 - {s_{71} \over s_{56}} \biggr) \,, \hskip 1 cm 
\\
&& B(3,4,6) = F^{{\rm 2m}\,h} (s_{71}, s_{567}, s_{234}, s_{56})
   \longrightarrow 
 F^{{\rm 1m}} (s_{71}, s_{567}, s_{56})
     + {1\over \e^2} (-s_{234})^{-\e} \nonumber 
\\
  && \hskip 6.5 cm \null
 - {1 \over 2 \e^2} {(-s_{234})^{-\e} (-s_{56})^{-\e} \over (-s_{71})^{-\e}}
 - \Li_2\biggl(1 - {s_{56} \over s_{71}} \biggr) \,,
\\
&& B(3,4,7) = F^{{\rm 2m}\,e}(s_{567}, s_{671},  s_{234}, s_{67}) 
   \longrightarrow 
  F^{{\rm 1m}}(s_{567}, s_{671}, s_{67}) 
  + {1\over \e^2} (-s_{234})^{-\e} \,,
\\
&& B(1,6,7) = F^{{\rm 1m}}(s_{23}, s_{34},  s_{234}) 
   \longrightarrow 
  F^{{\rm 0m}}(s_{23}, s_{34}) 
   + {1\over \e^2} (-s_{234})^{-\e} \,. \hskip 3 cm \null
\label{B167limit}
\end{eqnarray}
The two remaining integrals,
\begin{equation}
F^{{\rm 2m}\,h} (s_{12}, s_{234}, s_{34}, s_{567}), \hskip 1 cm 
F^{{\rm 2m}\,h} (s_{45}, s_{234}, s_{671}, s_{23}), 
\end{equation}
do not have particularly simple properties as 
$s_{234} \rightarrow 0$, although the polylogarithms 
do simplify to logarithms.

Using eqs.~(\ref{FirstFiveMulti})--(\ref{B167limit}), we may rewrite
\eqn{T234PoleInts} as
\begin{eqnarray}
&& A_{7;1}^{\NeqFour}(1^-,2^-,3^-,4^+,5^+,6^+, 7^+) \nonumber \\
 && \hskip 1 cm  \longrightarrow 
\cg \, (\mu^2)^\e A_4^{\rm tree} (2^-, 3^-, 4^+, (-K)^+)
 \, {i\over s_{234}} \,  A_5^{\rm tree}(K^-, 5^+, 6^+, 7^+, 1^-)
 \nonumber \\
&& \hskip 2.5 cm \times
\Biggl[   F^{\rm 1m}(s_{56}, s_{67}, s_{567})
        + F^{\rm 1m}(s_{67}, s_{71}, s_{671})
        + F^{\rm 1m}(s_{671}, s_{56}, s_{71}) \label{MultiLimitForm}\\
&& \hskip 2.9 cm \null
        + F^{\rm 1m}(s_{71}, s_{567},s_{56})
        + F^{\rm 1m}(s_{567}, s_{671}, s_{67})
        + 2 F^{\rm 0m}(s_{23},s_{34}) 
        +   {\cal R}_{234}
 \Biggr] \,,\hskip .3 cm \null  \nonumber 
\end{eqnarray}
where the remainder is 
\begin{eqnarray}
{\cal R}_{234} & = &
 {5 \over \e^2} (-s_{234})^{-\e}
 - {1 \over 2 \e^2} {(-s_{234})^{-\e} (-s_{71})^{-\e} \over (-s_{56})^{-\e}}
 - \Li_2\biggl(1 - {s_{71} \over s_{56}} \biggr) \nonumber \\
&& \hskip 0.3 cm \null 
 - {1 \over 2 \e^2} {(-s_{234})^{-\e} (-s_{56})^{-\e} \over (-s_{71})^{-\e}}
 - \Li_2\biggl(1 - {s_{56} \over s_{71}} \biggr) \nonumber\\
&& \hskip 0.3 cm \null 
 + 2 F^{{\rm 2m}\, h}(s_{12}, s_{234}, s_{34}, s_{567}) 
 + 2 F^{{\rm 2m}\, h}(s_{45}, s_{234}, s_{671}, s_{23})\,.
\label{Remainder234}
\end{eqnarray}
Comparing \eqn{MultiLimitForm} to \eqn{LoopFact234}, we may identify
\begin{equation}
A_{4;1}^{\NeqFour}(2^-,3^-,4^+, (-K)^+) = 
2 \, \cg \, (\mu^2)^\e \, A_{4}^{\tree} \, F^{\rm 0m}(s_{23}, s_{34}) \,,
\label{MPFourIdentify}
\end{equation}
in agreement with~\eqn{FourPointAmplitude}, and 
\begin{eqnarray}
 A_{5;1}^{\NeqFour}(K^-, 5^+, 6^+, 7^+, 1^-) &=&
\cg \, (\mu^2)^\e\, A_5^\tree \Biggl[ 
          F^{\rm 1m}(s_{56}, s_{67}, s_{567})
        + F^{\rm 1m}(s_{67}, s_{71}, s_{671})\nonumber \\
&& \hskip 2.3 cm \null
        + F^{\rm 1m}(s_{71}, s_{567},s_{56})
        + F^{\rm 1m}(s_{567}, s_{671}, s_{67}) \nonumber \hskip 1 cm \\
&& \hskip 2.3 cm \null
        + F^{\rm 1m}(s_{671}, s_{56}, s_{71}) 
                              \Biggr]\,, \hskip 1 cm
\label{MPFiveIdentify}
\end{eqnarray}
in agreement with with \eqn{FivePointAmplitude}. 
The remainder may also be identified with the factorization
function, 
\begin{equation}
\Fact_7(s_{234}; k_1, \ldots, k_7) = (\mu^2)^\e \, {\cal R}_{234}\,.
\end{equation}
The last equation requires use of a dilogarithm identity and
expansion through $\Ord(\e^0)$.  Combining this result 
with \eqns{MPFourIdentify}{MPFiveIdentify}, we thus find that the 
amplitude $A_{7;1}^{\NeqFour}(1^-,2^-,3^-,4^+,5^+,6^+,7^+)$ 
has precisely the expected factorization properties in the limit 
$s_{234} \rightarrow 0$.

The other channels work similarly. For all the amplitudes in
\sect{ResultsSection}, for each multi-particle limit the coefficients
of precisely eight integrals have poles in the channel of interest.
Five of the integrals belong with $A_{5;1}^{\NeqFour}$, one belongs with
$A_{4;1}^{\NeqFour}$, and the remaining two combine with the discontinuities
from the other integrals to form the factorization function.


\end{document}